\documentclass[twocolumn,tighten,trackchanges]{aastex61}
\usepackage{times}
\usepackage{natbib}
\usepackage{amssymb}
\usepackage{amsmath}
\usepackage{amstext}
\usepackage{xcolor}
\usepackage{float}
\usepackage[varg]{txfonts}  
\usepackage{microtype}      % improve the line-breaking
\usepackage{graphicx}
\usepackage{multirow}
\usepackage{booktabs}
\usepackage[normalem]{ulem}

\newcommand\afe{\ensuremath{[\alpha/\mathrm{Fe}]}}

\definecolor{orange}{RGB}{255,140,0}
\definecolor{darkred}{RGB}{139,0,0}
\definecolor{blueviolet}{RGB}{138,43,226}
\definecolor{forestgreen}{rgb}{0.13, 0.55, 0.13}
\definecolor{electricultramarine}{rgb}{0.25, 0.0, 1.0}

\received{}
\revised{}
\accepted{August 15th, 2018}

\shorttitle{The GeMS/GSAOI Galactic Globular Cluster Survey (G4CS), I.}
\shortauthors{Monty et al.}

\begin{document}

\title{The GeMS/GSAOI Galactic Globular Cluster Survey (G4CS) I: A Pilot Study of the stellar populations in NGC\,2298 and NGC\,3201}

\author[0000-0002-9225-5822]{Stephanie Monty}
\affiliation{Department of Physics and Astronomy, University of Victoria, Victoria, BC V8W 3P2, Canada}
\affiliation{Research School of Astronomy and Astrophysics, Australian National University, Canberra, ACT 2611, Australia}
\author[0000-0003-0350-7061]{Thomas H.~Puzia}
\affiliation{Institute of Astrophysics, Pontificia Universidad Cat\'olica de Chile, Av.~Vicu\~na Mackenna 4860, 782-0436 Macul, Santiago, Chile}
\author[0000-0002-5665-376X]{Bryan W.~Miller}
\affiliation{Gemini Observatory/AURA, Southern Operations Center, Casilla 603, La Serena, Chile}
\author[0000-0002-7272-9234]{Eleazar R.~Carrasco}
\affiliation{Gemini Observatory/AURA, Southern Operations Center, Casilla 603, La Serena, Chile}
\author[0000-0002-5652-6525]{Mirko Simunovic}
\altaffiliation{Gemini Science Fellow}
\affiliation{Gemini Observatory, 670 North A'ohoku Place, Hilo, HI 96720, USA}
\author{Mischa Schirmer}
\affiliation{Gemini Observatory/AURA, Southern Operations Center, Casilla 603, La Serena, Chile}
\affiliation{Max-Planck-Institut f\"ur Astronomie, K\"onigstuhl 17, D-69117 Heidelberg, Germany}
\author[0000-0001-6074-6830]{Peter B. Stetson}
\affiliation{Herzberg Astronomy and Astrophysics, National Research Council Canada, Victoria, BC V9E 2E7, Canada}
\author[0000-0001-5870-3735]{Santi Cassisi}
\affiliation{INAF - Astronomical Observatory of Abruzzo. Via M. Maggini, I-64100 Teramo Italy}
\author{Kim A. Venn}
\affiliation{Department of Physics and Astronomy, University of Victoria, Victoria, BC V8W 3P2, Canada}
\author[0000-0002-4442-5700]{Aaron Dotter}
\affiliation{Harvard-Smithsonian Center for Astrophysics, Cambridge, MA 02138, USA}
\author[0000-0002-5728-1427]{Paul Goudfrooij}
\affiliation{Space Telescope Science Institute, 3700 San Martin Drive, Baltimore, MD 21218, USA}
\author{Sibilla Perina}
\affiliation{INAF -- Astronomical Observatory of Torino, Via Osservatorio 20, I-10025 Pino Torinese, Italy}
\author[0000-0003-0454-5705]{Peter Pessev}
\affiliation{Gran Telescopio Canarias, Instituto de Astrof\'isica de Canarias, Universidad de La Laguna, Spain}
\author{Ata Sarajedini}
\affiliation{Department of Astronomy, University of Florida, 211 Bryant Space Science Center Gainesville, FL 32611 USA}
\author[0000-0003-3009-4928]{Matthew A. Taylor}
\altaffiliation{Gemini Science Fellow}
\affiliation{Gemini Observatory, 670 North A'ohoku Place, Hilo, HI 96720, USA}

\correspondingauthor{Stephanie Monty, Thomas H.~Puzia}
\email{montys@uvic.ca, tpuzia@astro.puc.cl}

\begin{abstract}
We present the first results from the GeMS/GSAOI Galactic Globular Cluster Survey (G4CS) of the Milky-Way globular clusters (GCs) NGC\,3201 and NGC\,2298.~Using the \textit{Gemini South Adaptive Optics Imager} (GSAOI), in tandem with the \textit{Gemini Multi-conjugate adaptive optics System} (GeMS) on the 8.1-meter Gemini-South telescope, we collected deep near-IR observations of both clusters, resolving their constituent stellar populations down to $K_s\simeq21$ Vega mag.~Point spread function (PSF) photometry was performed on the data using spatially-variable PSFs to generate $JHK_{s}$ photometric catalogues for both clusters.~These catalogues were combined with \textit{Hubble Space Telescope} (HST) data to augment the photometric wavelength coverage, yielding catalogues that span the near-ultraviolet (UV) to near-infrared (near-IR).~We then applied $0.14$ mas/year accurate proper-motion cleaning, differential-reddening corrections and chose to anchor our isochrones using the lower main-sequence knee (MSK) and the main-sequence turn-off (MSTO) prior to age determination.~As a result of the data quality, we found that the $K_{s}$ vs.~F606W~$\!-K_{s}$ and F336W vs.~F336W~$\!-K_{s}$ color-magnitude diagrams (CMDs) were the most diagnostically powerful.~We used these two color combinations to derive the stellar-population ages, distances and reddening values for both clusters.~Following isochrone-fitting using three different isochrone sets, we derived best-fit absolute ages of $12.2\pm0.5$ Gyr and $13.2\pm0.4$ Gyr for NGC\,3201 and NGC\,2298, respectively.~This was done using a weighted average over the two aforementioned color combinations, following a pseudo-$\chi^2$ determination of the best-fit isochrone set.~Our derived parameters are in good agreement with recent age determinations of the two clusters, with our constraints on the ages being or ranking among the most statistically robust.
\end{abstract}

\keywords{globular clusters: individual (NGC\,2298, NGC\,3201) --- galaxies: star clusters: general --- stars: low-mass --- infrared: stars --- ultraviolet: stars}

\section{Introduction} \label{sec:intro}
\subsection{Complexity of GC Stellar Populations}
The study of resolved globular clusters (GCs) provides unique leverage on some of the most pressing questions in astronomy, ranging from our understanding of stellar structure and evolution, to the details of star formation and chemical enrichment processes, to the formation histories of galaxies and the beginnings of structure formation in the universe \citep[e.g.][and references therein]{gra04, ren13, ren17, kru14, van17}.~The {\it Hubble Space Telescope UV Legacy Survey of Galactic Globular Clusters} \citep{pio15} and the {\it Hubble Space Telescope ACS Survey of Galactic Globular Clusters} \citep{sar07} obtained deep near-UV\footnote{F275W[UV] and F336W[$U$] WFC3 filters} and optical\footnote{F438W[$B$] WFC3, and F606W[$V$] and F814W[$I$] ACS filters} photometry for about 60 Milky Way GCs, enabling the creation of high-quality panchromatic color-magnitude diagrams (CMDs).~Analogous to the theoretical stellar evolution-tracing Hertzsprung-Russell (HR) diagrams, deep CMDs can be used to test theories of stellar evolution and star cluster formation.~Individual GCs constitute the best available approximations to simple stellar populations, being composed of coeval stars with virtually the same chemical composition.~As such, they frequently serve as calibrators for stellar population synthesis models which can then be used to study distant, unresolved stellar systems.~The deep CMDs that arose from the Hubble Space Telescope (HST) surveys have produced a range of important benchmark measurements and several unexpected findings.~For instance, it was found that a number of Galactic GCs harbour multiple stellar populations \citep[e.g.][]{mil08, pio15}, a result with far reaching consequences for our understanding of their formation \citep[e.g.][]{ren15} and with close ties to star formation physics.~The second prominent result was the determination of GC relative ages with sufficient precision to directly probe the formation and evolution of the galactic halo \citep[e.g.][]{mar09, dot10, dot11, lea13}.~These results suggest stark differences between the early formation history of the inner and outer halo, providing further evidence that the evolution of the outer halo involved the disruption and accretion of dwarf satellite galaxies \citep[e.g.][]{sea78, fre02, hel08, ive12}.

\subsection{The Role of Near-Infrared Observations}
Near-infrared (near-IR) CMDs offer significant advantages when compared to pure UV-optical CMDs, such as:~{\it i}) near-IR colors are much less affected by differential reddening than optical colors and~{\it ii}) near-IR isochrones exhibit a very characteristic S-shaped main sequence (MS) which can be used for accurate, {\it absolute} age determinations.~This characteristic shape is attributed to low-mass MS stars, which show a well-defined (faint-ward) bending, due to collision-induced and roto-translational absorption of hydrogen molecules (H$_2$-H$_2$ and other molecular pairs, such as: H$_2$-He, H$_2$-CH$_4$, etc.)~at near-IR wavelengths \citep[also referred to as the MS ``knee'' or MSK;][]{bor01, bor02, gus09, fro10, ric12}.~The resultant difference in color and luminosity between the MS turn-off (MSTO) point and the lower MSK depends predominantly on metallicity, while the luminosity and color of the MSK are virtually independent of age at any given metallicity \citep{cal09}.~Additionally, because the convective motions in the stellar atmospheres are mostly adiabatic at low stellar masses \citep{sau08}, the MSK is anchored in a regime of well understood stellar structure and evolution.~Therefore, by exploiting the metallicity-dependence-dominated MSTO-MSK distance, the age-independence of the MSK at fixed metallicity and its well understood physics, the observed MSTO-MSK color and luminosity difference form a robust {\it absolute} theoretical reference frame which allows for the measurement of absolute GC ages with uncertainties of about 1~Gyr.~This remarkable property of near-IR CMDs was recently exploited by \cite{mas16}, \cite{cor16}, and \cite{sara16} using various isochrone fitting techniques.~Finally, because near-IR CMDs contain the two aforementioned MS reference points, i.e.\ the MSTO and MSK, the age information extracted from their analysis becomes independent of cluster distance and screen-type reddening effects.

\subsection{The GeMS/GSAOI Galactic Globular Cluster Survey}
In this paper we describe the first results from the {\it GeMS/GSAOI Galactic Globular Cluster Survey} (G4CS).~The G4CS is conducting a homogeneous, deep near-IR CMD study of Milky Way GCs covering a broad spectral energy distribution (SED) baseline in order to: {\it i}) investigate the morphology of near-IR CMDs, particularly at low stellar masses, {\it ii}) determine the internal consistency and calibration of model CMD predictions for various near-IR+optical+near-UV filter combinations, thereby testing the influence of variations in the molecular absorption bands on isochrones using information from publicly-available high-resolution spectroscopy \citep[e.g.][]{car09, car10},~{\it iii}) derive absolute GC ages with accuracies of about 1~Gyr,~and {\it iv}) quantify and characterize binary fractions and blue-straggler star (BSS) populations.

\begin{figure}[t!]
   \centering
   \includegraphics[width=\linewidth]{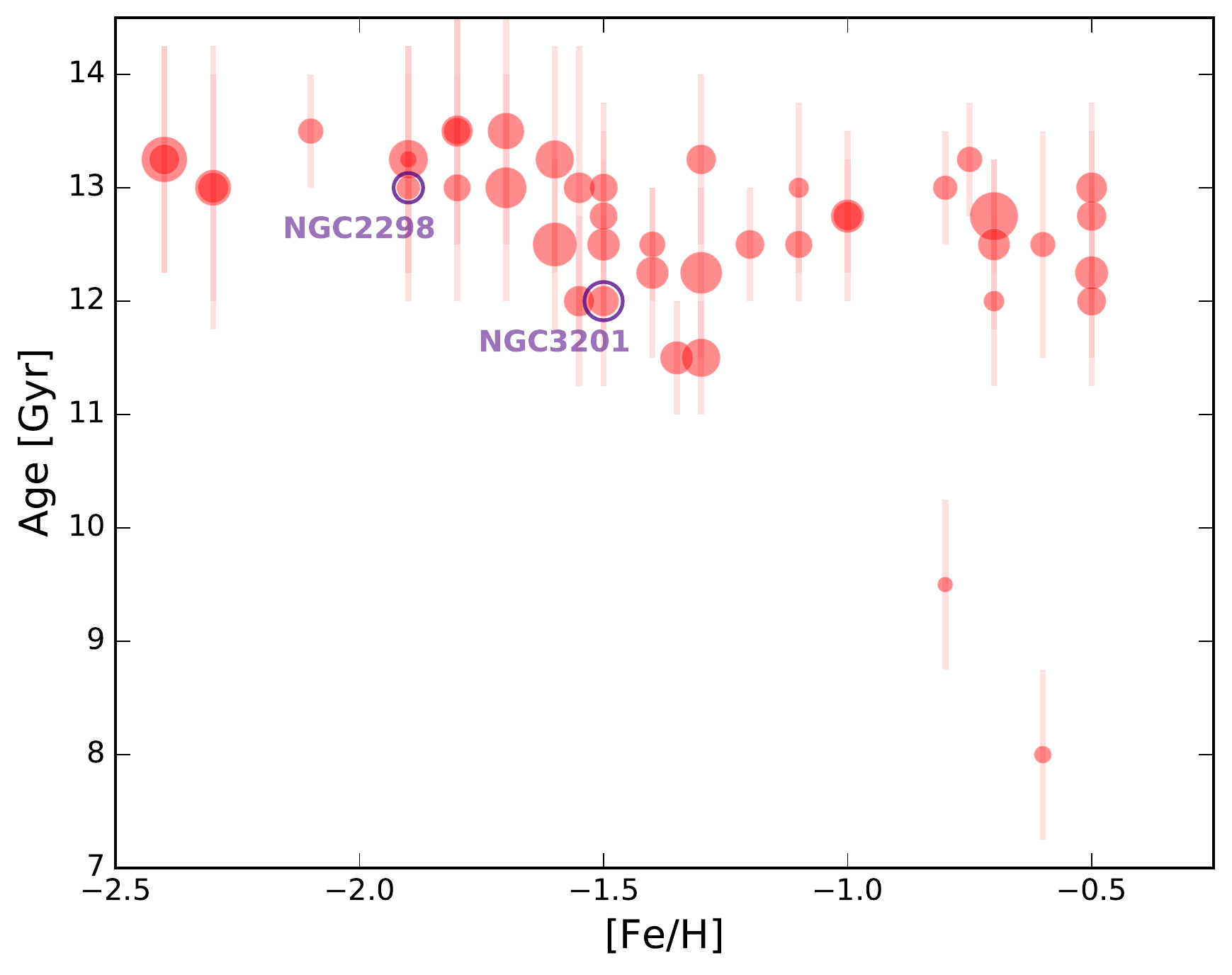}
   \caption{The age-metallicity relationship for our full G4CS target sample with available data from \cite{dot10}.~The two GCs from this study are circled and correspondingly labeled.~The symbol size scales with the $V$-band integrated luminosity; larger symbols correspond to brighter GCs.~Note that the whole sample covers the transition region between the flat and steep age-metallicity relation.~Utilizing our excellent age/metallicity resolution, we aim to clearly resolve the transition between these two GC sub-populations.}
   \label{fig:sample}
\end{figure}

The G4CS target list was compiled through selecting GCs included in the recent HST surveys of Galactic GCs, to ensure that deep near-UV+optical photometry is available \citep{sar07,pio15}, with the additional restriction that the GCs lay no further than 23~kpc to ensure that the MSK can be detected in a reasonable amount of time with GeMS/GSAOI at the 8.1-meter Gemini-South telescope.~In order to study as broad a parameter space as possible, the target sample covers a range in GC ages ($t\!\simeq\!8.0\!-\!13.5$~Gyr), metallicities ($[{\rm Z/H}]\!\simeq\!-2.37$ to $-0.32$~dex), and masses (${\cal M}_{GC}\!\simeq\!10^{3.75}\!-\!10^{6.05}M_{\odot}$, see also Fig.~\ref{fig:sample}).

This paper presents the results for the first two GCs observed from the sample: NGC\,2298 and NGC\,3201.~These two GCs were selected, in part, due to little indication of multiple stellar populations in their near-UV/optical CMDs \citep{pio15}, making them two of the few GCs that may most closely resemble simple stellar populations.~Both clusters have HBs that could be influenced by the ``2nd parameter'' effect and NGC\,3201 may be about a Gyr younger than the oldest GCs \citep{mil14}.~This implies that NGC\,3201 is a candidate for having been formed in a dwarf galaxy and later accreted into the Milky Way halo \citep{mac04}.~Indeed, evidence of extra-tidal stars has been found around both clusters \citep{balb11,ang16}.~Both clusters also experience relatively high reddening ($E_{B-V} \simeq 0.2$\,mag), making it advantageous to study them in the near-IR.~Additionally, NGC\,3201 may be considered a control cluster since it has been used by both \cite{cal09} and \cite{bon10} to demonstrate the MSTO-MSK method using VLT MAD and HST, respectively.~Therefore, being able to compare our analysis to these earlier results will allow us test the efficacy of our methods.~The properties of the two GCs studied thus far are presented in Table~\ref{tab:gcprop}.

\begin{deluxetable}{lrr}
\tablecaption{Fundamental GC parameters. \label{tab:gcprop}}
\tablehead{
\colhead{Parameter}   & \colhead{NGC 3201}   & \colhead{NGC 2298}
}
\startdata
Distance [kpc]        & 4.9                  & 10.8\\
$(m-M)_V$             & 14.2                 & 15.6\\
$E_{B-V}$           & 0.24                 & 0.14\\
$r_h$ [pc]            & 4.4                  & 3.1\\
$M_{V}$             & $-7.45$              & $-6.31$\\
$[{\rm Fe/H}]$        & $-1.59$              & $-1.92$\\
\enddata
\tablecomments{All values are taken from the Milky Way GC McMaster Catalog \cite{har10}. Note that the distance moduli and absolute magnitudes are not corrected for reddening, while the distance has been corrected.}
\end{deluxetable}

\begin{figure*}[t]
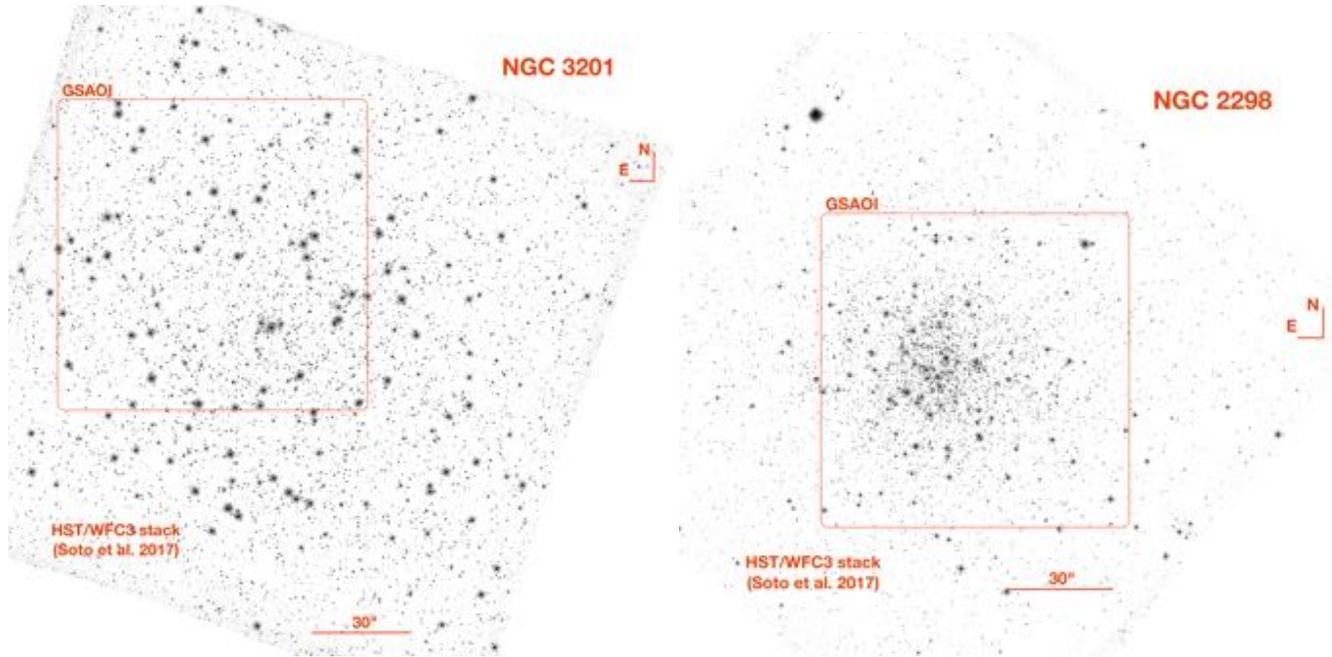

   \centering
   \includegraphics[width=0.49\linewidth]{Figure2a.pdf}
   \includegraphics[width=0.49\linewidth]{Figure2b.pdf}
   \caption{HST/WFC3 stacked UV+optical (F275W+F336W+F438W) greyscale images of NGC\,3201 ({\it left}) and NGC\,2298 ({\it right}) from \citet{sot17}.~The red squares represent the area covered by our GeMS/GSAOI observations.}
   \label{fig:hstfields}
\end{figure*}

This paper is organized as follows: the observations and data reduction are presented in Section~\ref{sec:obs}, in Section~\ref{sec:anls} we describe the various experiments undertaken to obtain the best PSF photometry given the characteristics of our observations, describe zero-point calibration, proper-motion cleaning and differential reddening corrections,  Section~\ref{sec:results} describes our methodology and results of isochrone-fitting and, finally, Section~\ref{sec:sum} concludes with a summary of the work.

%%%%%%%%%%%%
\section{Observations and Data Reduction} \label{sec:obs}

\subsection{NGC\, 3201 and NGC\, 2298 data acquisition}\label{sec:dataacq}
The observations were conducted using the {\it Gemini South Adaptive Optics Imager} \citep[GSAOI; for details see][]{mcg04, car12}.~GSAOI is fed by the {\it Gemini Multi-conjugate adaptive optics System} \citep[GeMS;][]{rig14, nei14a, nei14b}, a facility Adaptive Optics (AO) system mounted on the Gemini-South telescope located on Cerro Pach\'on in Chile.~GeMS uses five lasers guide stars (LGS) to correct for atmospheric distortion and up to three Natural Guide Stars (NGS) brighter than $R\!=\!15.5$ mag to compensate for tip-tilt and plate scale variations over the $2\arcmin$-diameter~field-of-view (FoV) of the AO bench unit \citep[CANOPUS,][]{rig14}.~GSAOI is a near-IR AO camera and companion instrument to GeMS, which receives and images the corrected near-IR light output from GeMS.~Together they deliver near-diffraction limited images in the wavelength interval of 0.9 -- 2.4\,$\micron$.~The GSAOI focal plane consists of a $2 \times 2$ mosaic of Rockwell HAWAII-2RG $2048 \times 2048$ detector arrays and provides a FoV of 85~$\times$ 85~arcsec$^2$~on the sky with a scale of $0\farcs02$ per pixel and gaps of $\sim\!2\farcs5$ between arrays.

The central regions of the two target Galactic GCs, NGC\,3201 and NGC\,2298, were imaged through the $z$, $J$, $H$ and $K_{s}$ standard GSAOI filters\footnote{Filter transmission curves are available at \url{http://www.gemini.edu//sciops/instruments/gsaoi/instrument-description/filters}.}.~Figure~\ref{fig:hstfields} shows the area covered by our GeMS/GSAOI observations for NGC\,3201 (\textit{left}) and NGC\,2298 (\textit{right}) GCs overlaid on top of HST/WFC3 near-UV+optical stacked images from \cite{sot17}.~All images were acquired in photometric conditions and with a variable natural seeing.

The observational strategy consisted of obtaining deep images in all four bands to cover a wide near-IR SED baseline and to study the CMD morphology down to the MSK.~This strategy allowed us to reach faint magnitudes (see Sect.~\ref{sec:anls}), while sacrificing the bright end of the luminosity function as all stars brighter than $K_s\simeq 12$\,mag were saturated.~Nine, 30-second exposures were obtained for each GC in each filter, using an offset of 5\arcsec\ between individual exposures, following a $3\times3$ dither-pattern to fill the gaps between the arrays.~A few sub-exposures suffered from the effect of the variable seeing present during the observations, reducing the effective exposure time in some filters.~This was the case for NGC\,3201, where one exposure from both the $J$ and $H$ filters had to be removed from the dataset.~The entire dataset obtained with the $z$ filter, for both GCs, was unusable due to poor image quality (the average AO-corrected full-width at half maximum, FWHM, of the stellar PSF was larger than $0\farcs3$).~Therefore, the $z$-band datasets are not considered in the subsequent analysis.~Table~\ref{tab:fwhm} summarizes the values of natural seeing measured by the differential image motion monitor (DIMM) at Cerro Pach\'on, the total exposure time, average airmass of the observations, average resolution derived from stars over the field (AO-corrected FWHM) and the average Strehl ratios estimated from the observations for the two GCs.

\begin{figure*}[t!]
   \centering
   \includegraphics[width=0.8\linewidth]{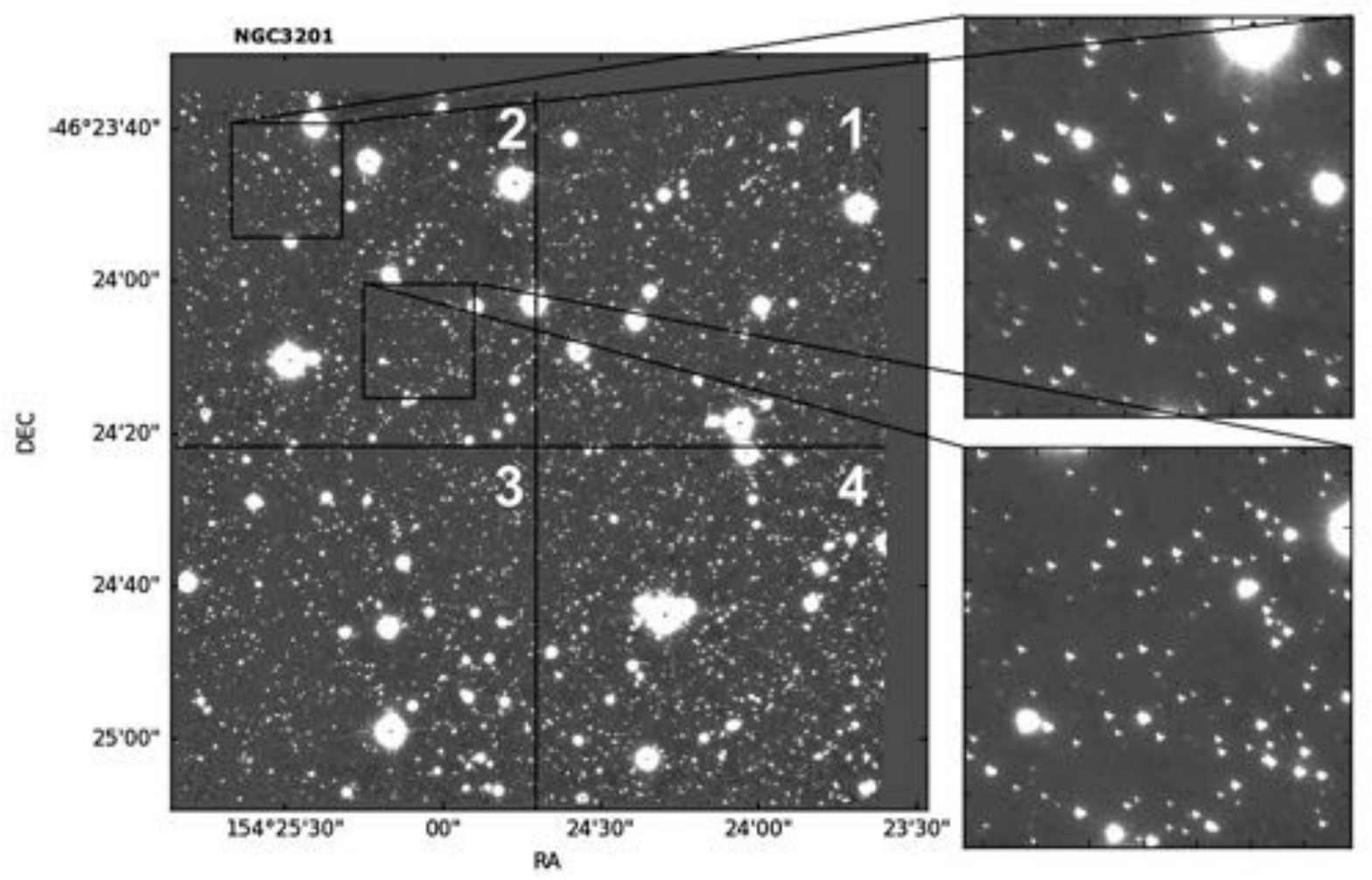}
   \includegraphics[width=0.8\linewidth]{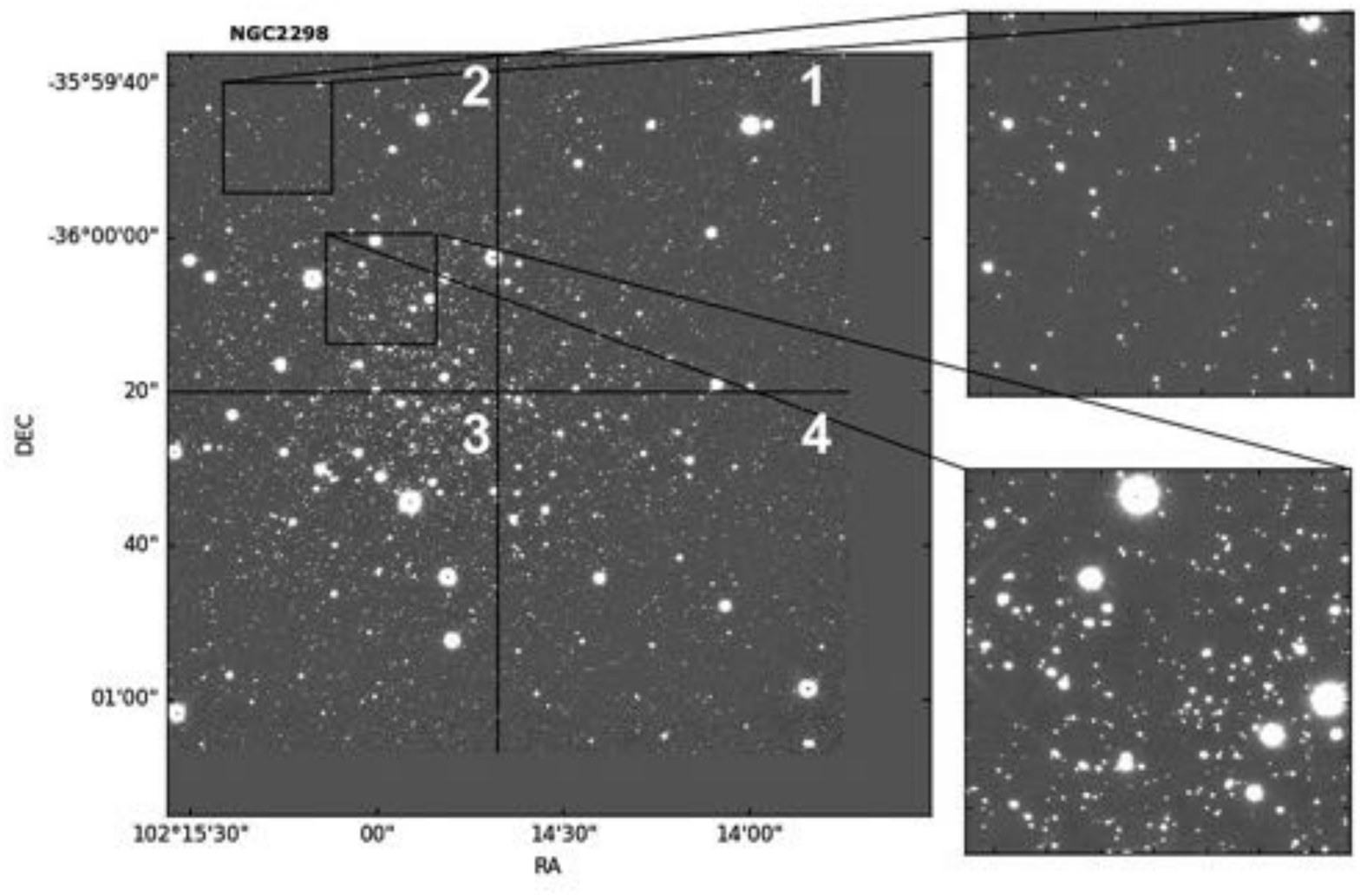}
   \caption{GeMS/GSAOI $K_s$-band co-added mosaic images of NGC\,3201 (top) and NGC\,2298 (bottom).~The insets show a region of array 2 with a sampling of the worst (\textit{top} zoom-in image) and best PSFs (\textit{bottom} zoom-in image).~The labels 1-4 indicate the numbering of each array.}
   \label{fig:fields}
\end{figure*}

\begin{deluxetable}{clccccc}[h!]
\tabletypesize{\scriptsize}
\tablecaption{Observing Log. \label{tab:fwhm}}
\tablehead{
\colhead{Cluster} & \colhead{Filter} & \colhead{Exp.time} & \colhead{Airmass} & \colhead{Natural} & \colhead{AO} & \colhead{Strehl} \\
\colhead{} & \colhead{} & \colhead{[s]} & \colhead{} & \colhead{Seeing} & \colhead{FWHM} & \colhead{[\%]}}
\startdata
             & $J$  & $8 \times 30$ & 1.062 &             & 0$\farcs$14 & 3  \\
   NGC\,3201 & $H$  & $8 \times 30$ & 1.073 & 0$\farcs$72 & 0$\farcs$12 & 8  \\
             & $K_s$& $9 \times 30$ & 1.085 &             & 0$\farcs$11 & 15 \\ \hline
             & $J$  & $9 \times 30$ & 1.124 &             & 0$\farcs$10 & 6  \\
   NGC\,2298 & $H$  & $9 \times 30$ & 1.149 & 0$\farcs$83 & 0$\farcs$08 & 10  \\
             & $K_s$& $9 \times 30$ & 1.177 &             & 0$\farcs$08 & 24 \\
\enddata
\tablecomments{NGC\,3201 and NGC\,2298 were observed during the nights of January 12--13, 2014 (program ID: GS-2013B-Q-61) and February 12--13, 2014 (program ID: GS-2014A-Q-13), respectively.}
\end{deluxetable}

\subsection{Laser Guide Star Wavefront Sensor Misalignment Problem}\label{sec:lgswfsm}
\begin{figure*}[t!]
   \centering
   \includegraphics[width=\linewidth]{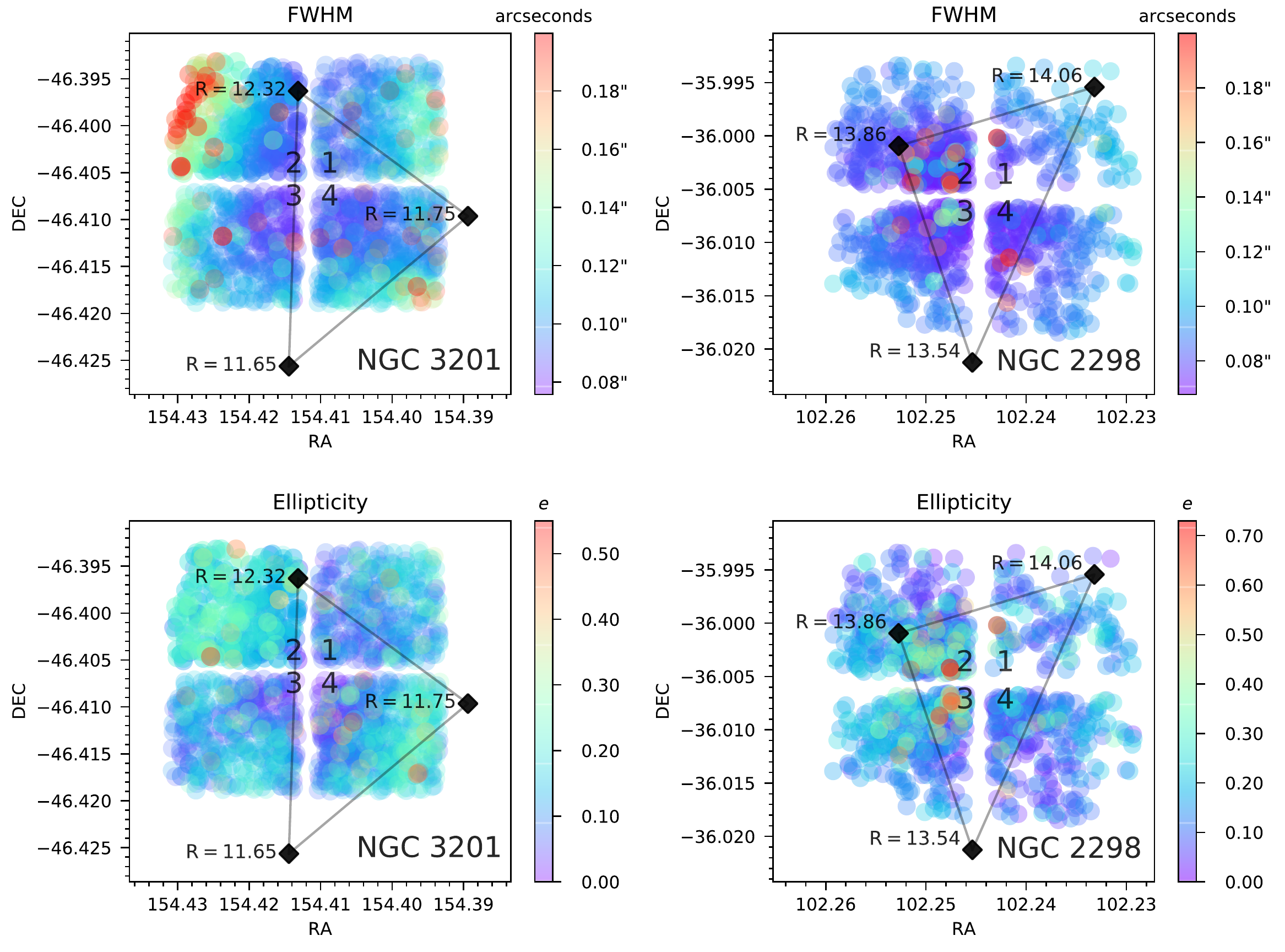}
   \caption{FWHM (\textit{top panels}) and ellipticity maps (\textit{bottom panels}) of the NGC\,3201 fields (\textit{left panels}) and NGC\,2298 (\textit{right panels}) in the $K_s$ band, created from a sample of $\sim\!500$ stars in the corresponding mosaic, co-added images.~The NGS constellation is indicated by the black diamonds and lines, with the $R$-band magnitudes of the stars labeled correspondingly.~1-4 at the center of the FoV indicate the array numbering.}
   \label{fig:fwhmell}
\end{figure*}

The point spread functions (PSFs) of the stars are uncharacteristically elongated and distorted in both datasets.~The PSF distortions are more pronounced at the edges of the images and they are much larger than the variations expected in typical multi-conjugated adaptive optics (MCAO) images.~In typical GeMS/GSAOI images the PSF is more or less uniform across the GSAOI FoV, with a variation at the level of 8\% to 20\% \citep[e.g.][]{rig14, nei14a, nei14b, dalessandro16}.~The amount of variation depends on the selected NGS constellation, the number of NGSs used, the observing conditions (mainly seeing), the laser photon return, and other parameters \citep{rig14, nei14a}. Even for fields where only one NGS is available, the PSF is fairly uniform across the GSAOI FoV \citep[see][]{sch15}.~The source of the distortions detected in the images was traced to a temporary misalignment in the laser guide star wavefront sensor (LGSWFS) located inside the AO bench unit {\sc CANOPUS}\footnote{The report with the detailed discussion about technical aspects of the LGSWFS misalignment problem and subsequent solution \citep{sivo14} is available upon specific request from the Gemini Observatory staff.}.~The LGSWFS misalignment introduced additional distortions to the PSF, affecting the PSF performance of all the GeMS/GSAOI images observed between November 2013 and June 2014, during which the data for our two GCs were collected.~The NGC\,3201 dataset suffered more from the LGSWFS misalignment than the NGC\,2298 dataset, as the NGC\,2298 data was collected after the implementation of a software fix for the misalignment during the February 2014 observing block.~The extent of the effect of the LGSWFS misalignment is highlighted in the insets of Figure~\ref{fig:fields}, showing the $K_s$-band GeMS/GSAOI co-added mosaic images of NGC\,3201 (top) and NGC\,2298 (bottom).~The insets in Figure~\ref{fig:fields} correspond to a region located inside array 2, the array associated with the worst degradation of the PSF, with a sample of the worst and best PSFs given (top and bottom insets for NGC\, 3201 and NGC\, 2298 respectively).~In the NGC\,2298 dataset the PSF is more uniform across the FoV and the shape is less affected by the LGSWFS misalignment, while in NGC\,3201 the shape of the PSF is more variable across the FoV and the objects show elongations that mimic a ``mouse-ear'' form.~Figure~\ref{fig:fwhmell} shows the FWHM and ellipticity variations across the observed fields in both GCs.~In this figure one can see the effect of the LGSWFS misalignment along the edges of the four arrays, particularly, as mentioned, in the upper left-hand corner of array 2 for NGC\,3201.~The solution developed during the course of this analysis to compensate for the image degradation is presented in Section~\ref{sec:anls}.

\subsection{Data Reduction} \label{sec:datared}

The data were reduced using the {\sc Theli} software package \citep{sch13, erb05} following standard near-IR data reduction steps, beginning with the application of a non-linearity correction to all raw images.~Flat-field calibrations were performed using lamp-on (bright) and lamp-off (dark) dome flats, in order to create median combined, master, dark and bright flats.~The master dark flat was then subtracted from the master bright flat, after which  the resulting frame was normalized to unity.~Science frames were then subsequently flat-fielded and the gain differences between different arrays adjusted based on the master flats.~The background in each filter was then constructed from the dithered science frames (eight to nine images depending on the object and filter used) using a two-pass background model according to the following recipe.~\textit{i)} The flat-fielded science exposures were combined by their median without any masking, to remove the bulk of the sky signal.~\textit{ii)} SExtractor \citep{ber96} was run to identify and mask all objects with a detection threshold of 1.5 sigma above the sky and with a minimum area of 5 pixels.~To minimize the effect of fainter halos and avoid dark halos in the final co-added images, the isophotal ellipses for masking were enlarged using a \textit{Mask Expansion Factor} between 6 to 10, minimizing the dark halo effect in the co-added images and in the photometric precision \citep[see][for a discussion]{sch15}.~\textit{iii)} The resulting background models were then rescaled to match individual background levels of the original images and then subtracted.

The astrometric calibration and distortion correction were then derived using the program SCAMP \citep{ber06} within {\sc Theli}.~We constructed the reference catalog from archival HST/ACS F814W ($I$-band) images, as these images have precise astrometric and distortion corrections and fully overlap with the GSAOI fields of NGC\,3201 and NGC\,2298 (see Fig.~\ref{fig:hstfields}).~For each GC, a common astrometric solution was then derived to ensure that all filters were resampled using common WCS and pixel positions.~Prior to creating the mosaic science frames, sky-background subtraction was performed on individual images to homogenize the zero background level across the four GSAOI arrays.~The resulting images where then co-added using the application SWarp \citep{ber10} within {\sc Theli}.~Figure \ref{fig:fields} shows the $K_s$-band final co-added images for NGC\,3201 (\textit{top}) and NCG\,2298 (\textit{bottom}).

%%%%%%%%%%%%%%
\section{Analysis}
\label{sec:anls}
\subsection{Photometry}

Final PSF-fitting photometry was performed using \texttt{DAOPHOT-IV/ALLSTAR} \citep{stet94} using a cubically variable PSF, on the mosaicked, stacked image associated with the $J, H$ and $K_{s}$ bands.~The StarFinder \citep{dio00} and \texttt{DAOPHOT-II} \citep{stet87} packages were also tested prior to performing the final analysis, but \texttt{DAOPHOT-IV} proved to be the best option for continued analysis.~A cubically variable PSF model, only available with \texttt{DAOPHOT-IV}, was required to map the spatial variations in the PSF from CANOPUS which were compounded by the LGSWFS misalignment problems (see Sect.~\ref{sec:lgswfsm}).~Stellar positions were known \textit{a priori} via the HST/ACS catalogues for both clusters \citep{sar07}, which are deeper than our own, allowing for forced photometry to be performed.~Approximately $100\!-\!120$ stars were chosen per chip to model the PSF, totaling around 400 over the whole field in order to build a global PSF.~A Moffat function of $\beta\!=\!1.5$ was chosen as the best approximation from which to generate the PSF via a $\chi^{2}$ test using the output of \texttt{DAOPHOT-IV}.~Experiments were also conducted varying the number of PSF stars chosen to build the model PSF, but no improvement was seen in the final CMD or in the photometric error distributions, when more than around 400 stars were chosen over the entire field.

\begin{figure*}[t!]
   \centering
   \includegraphics[width=\linewidth]{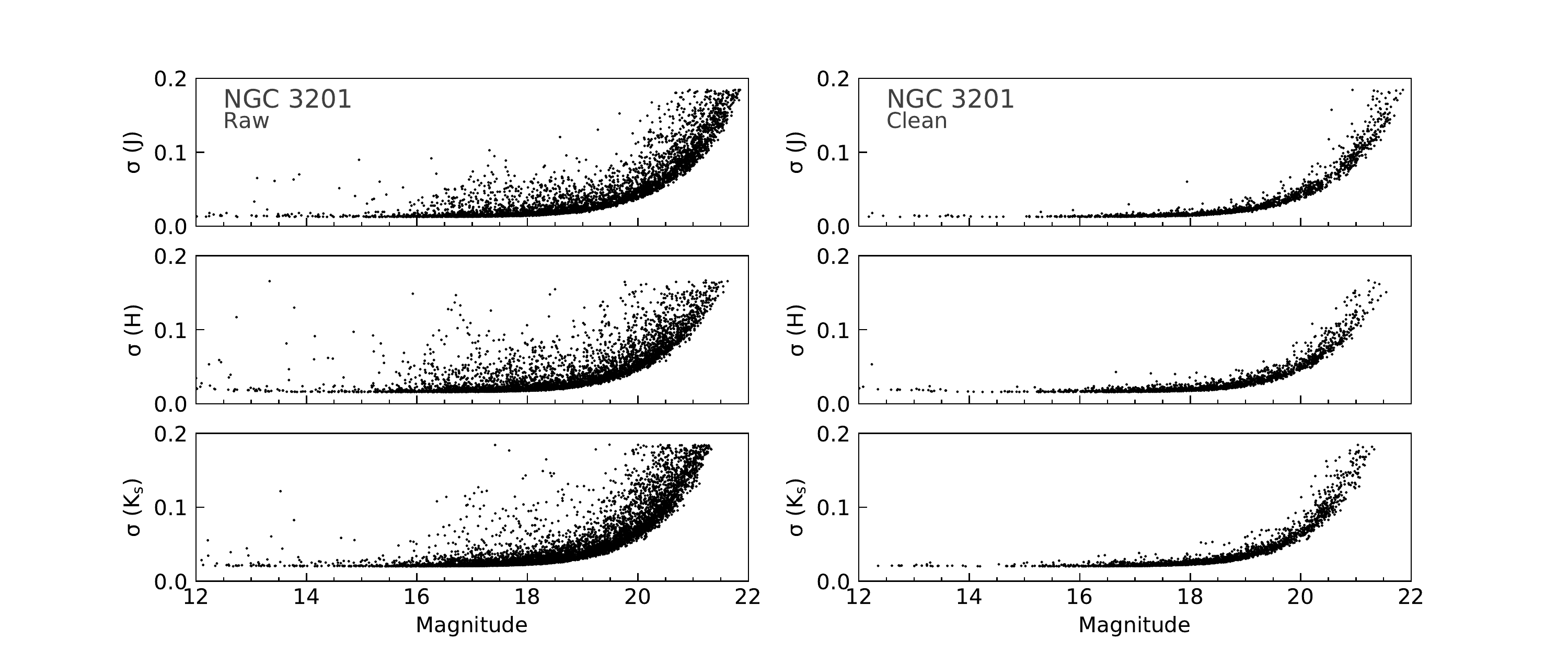}
    \caption{Photometric errors in units of magnitude as a function of magnitude in the $JHK_s$ filters ({\it top to bottom panel}), for the cluster NGC\,3201 following $1\sigma$-clipping, before (\textit{left panel}) and after (\textit{right panel}) proper-motion cleaning.}
    \label{fig:magerr}
\end{figure*}

While other papers have performed a chip-by-chip photometric approach, generating chip-specific PSFs \citep{sara15, tur15, tur17}, an exploration into any improvement made by performing photometry in this way proved unfruitful.~The decision to perform the final photometry on the mosaicked, stacked images was based primarily on the observed temporal variability and quality of the PSF between individual images.~In general, irregularities in the PSF in individual images were seen to average out in the stacked image, leading to a more symmetric and uniform PSF in the final stacked image.~Additionally, model PSFs were found to be successfully generated more often using the symmetric PSFs found in the stacked images, qualified as smaller fitting residuals in the final PSF-subtracted image.~This decision was further supported following an experiment directly comparing magnitudes measured using our own method against those of the chip-by-chip approach.~Photometry was performed, chip-by-chip, exposure-by-exposure for the NGC\,3201 data taken with the $K_{s}$ filter, after which an average magnitude was calculated for each data point and compared against the mosaic, stacked measurement using the global PSF.~On average the two measurements deviated by only $\sim\!1\%$, while the process of chip-by-chip PSF generation often presented substantial difficulties due to a lack of acceptable stars from which to generate the model PSF.~As a precaution, all stars laying in the chip gaps were removed from the final photometric catalogue to ensure no instances of repeated measurements due to a star appearing in more than one chip, a possible result of the dithering pattern.~Photometric uncertainties, as provided by \texttt{ALLSTAR} and presented as a function of magnitude can be seen for NGC\,3201 in Figure~\ref{fig:magerr}.~Note that the photometric catalogues for both clusters were cleaned via  $1\sigma$ clipping above the mean photometric error for all magnitudes, resulting in the truncated shape of the curves.~Figure~\ref{fig:magerr} also shows the improvement to the overall photometric quality of the catalogues following proper motion cleaning as described in Section~\ref{sec:PM}, with the initial $1\sigma$-clipped catalogue of NGC\,3201 being shown before (left) and after (right) proper motion cleaning. 

\subsection{Photometric Calibration}
\label{sec:photcal}
A limitation encountered in deriving the photometric zero-points associated with the GSAOI images, was the high resolution provided by the system.~Due to the absence of standard star catalogs derived using a similar resolution, zero-point calibration using standard stars from the 2MASS Point Source Catalog \citep{skrut06} was not possible, as most of the bright stars suitable for photometric calibration in the GSAOI images were saturated.~Therefore, the photometric zero-point calibration was performed using a two-step process involving intermediate standards.~Stars from the 2MASS catalog were identified and used to calibrate wider-field, but lower resolution, images that included the GSAOI fields.~Then, well isolated stars in the GSAOI frames were identified in the wide-field images and used as secondary standards to calibrate the GSAOI images to the 2MASS system.~In the case of NGC\,3201 archival ESO-VLT/{\sc Hawk-I} $JHK_s$ images were used for the calibration.~No archival data was available NGC\,2298 so we obtained $JHK_s$ imaging using {\sc Flamingos-2} at Gemini South during the night of August 6 -- 7, 2016 as part of program GS-2016B-Q-56.~Both datasets were processed with {\sc Theli}, using a method similar to the one described in Section~\ref{sec:datared}.~Instrumental magnitudes for non-saturated stars within the calibration data sets were obtained using aperture photometry, performed using different radii with the {\sc IRAF}\footnote{IRAF is distributed by the National Optical Astronomy Observatories (NOAO), which is operated by the Association of Universities for Research in Astronomy (AURA), Inc. under cooperative agreement with the National Science Foundation.} task {\sc phot} inside the {\sc daophot} package.~The measured magnitudes were then corrected by the aperture variation prior to calibration to the 2MASS photometric system.~The GSAOI images were then photometrically calibrated using the same approach, aperture photometry was performed using a set of well isolated non-saturated stars found in common with the {\sc Hawk-I} and {\sc Flamingos-2} images, from which the GSAOI frames were calibrated.~For all bands, a zeroth-order polynomial (constant) fitting to the difference in magnitude between GSAOI and {\sc Hawk-I}/{\sc Flamingos-2} data was used to derive the final zero-points.~Table~\ref{tab:zp} shows the photometric zero-points determined for the $JHK_s$ filters.~Note that unless otherwise stated, all magnitudes presented in this paper are in the Vega magnitude system.

\begin{deluxetable}{lcc}
\tablewidth{0.75\textwidth}
\tablecaption{2MASS zero-points. \label{tab:zp}}
\tablehead{
\colhead{Filter}    & \colhead{NGC 3201}    & \colhead{NGC 2298}
}
\startdata
$J$      & $26.403 \pm 0.017$  & $25.098 \pm 0.012$ \\
$H$      & $26.476 \pm 0.02\phn$   & $25.285 \pm 0.023$ \\
$K_{s}$  & $25.696 \pm 0.017$ & $24.443 \pm 0.034$ \\
\enddata
\tablecomments{Vega mag photometric zero-points, determined using 2MASS stars in HAWK-I images for NGC\,3201 and in imaging data from Flamingos-2 for NGC\,2298, as described in Section \ref{sec:photcal}.}
\end{deluxetable}

\subsection{Proper Motion Cleaning}\label{sec:PM}
In order to make our CMDs as uncontaminated as possible, we performed relative proper-motion (PM) measurements of all objects in both GCs, using the PM vector distributions to select high-likelihood cluster members.~The GSAOI ($x,y$) position catalogs that were used correspond to the centroid positions obtained from the \texttt{DAOPHOT} PSF-fitting photometry, performed chip-to-chip and exposure-by-exposure, contrary to our final photometric method used for isochrone fitting.~As the proper motion cleaning necessitates more accurate centroiding, photometric precision was sacrificed, again quantified through the fitting residuals.~We chose to use the $K_s$-band images to calculate relative proper-motions, as this filter provided the best FWHM (see Table~\ref{tab:fwhm}).~The counterpart catalogs from the HST/ACS GC Survey provided the ACS camera ($x,y$) coordinates, which are distortion-free and, hence, can be used as the reference frame and as first-epoch positions.

\begin{figure}[t]
   \centering
   \includegraphics[width=1\linewidth]{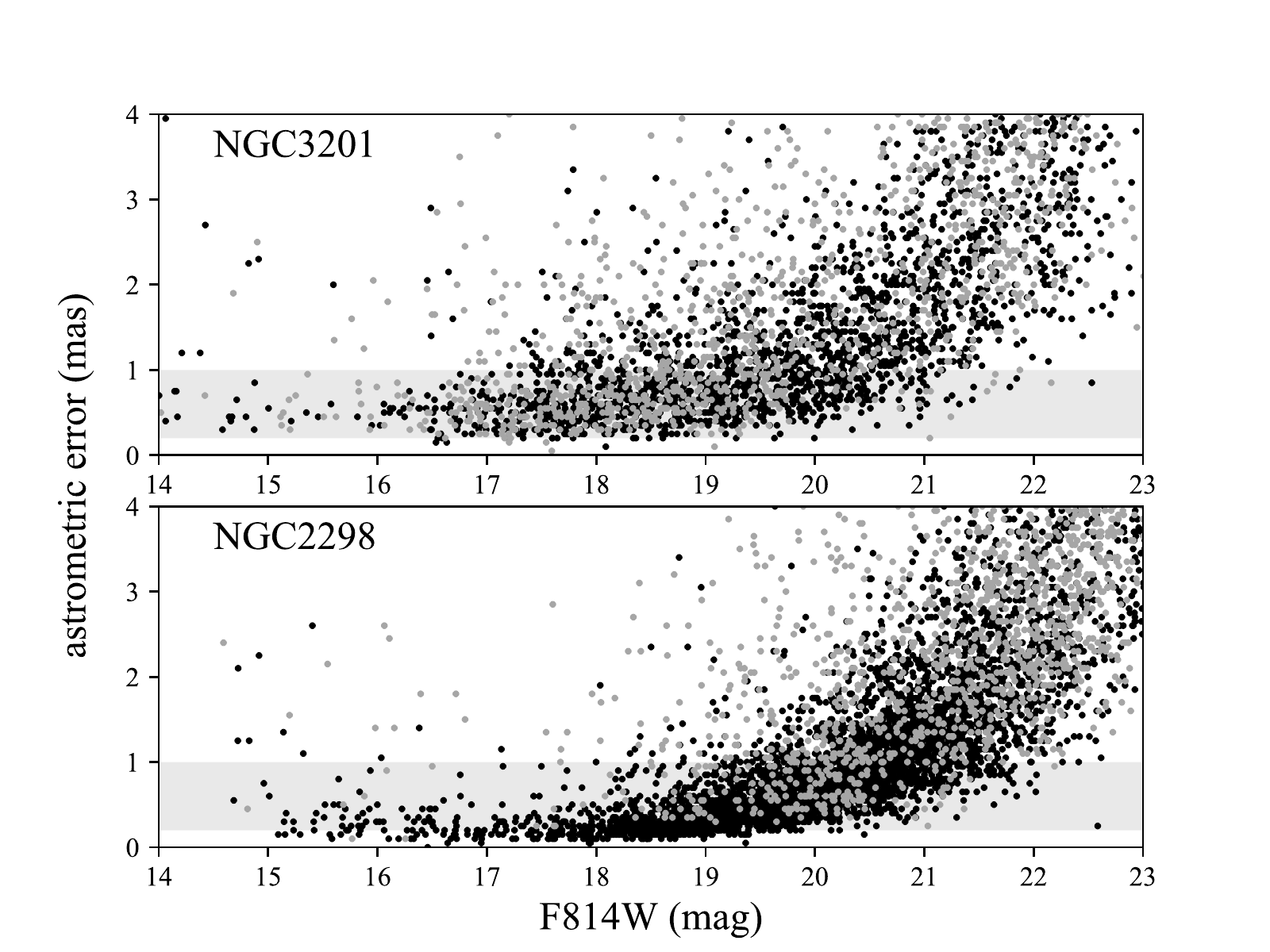}
   \caption{Astrometric error as a function of F814W magnitude for NGC\,3201 ({\it top panel}) and NGC\,2298 ({\it bottom panel}).~The grey points correspond to the rejected stars in the vector point diagram selection (see Figure~\ref{fig:pm_selection}).~The shaded regions mark the 0.2$-$1.0 mas range for the precision expected based on similar studies in the literature \citep{nei14b,mas16,dalessandro16}.~Note the comparatively worse performance in the NGC\,3201 data due to the PSF degradation explained in Sec.~\ref{sec:lgswfsm}.}
   \label{fig:pm_error}
\end{figure}

\begin{figure*}[t]
   \centering
   \includegraphics[width=0.49\linewidth]{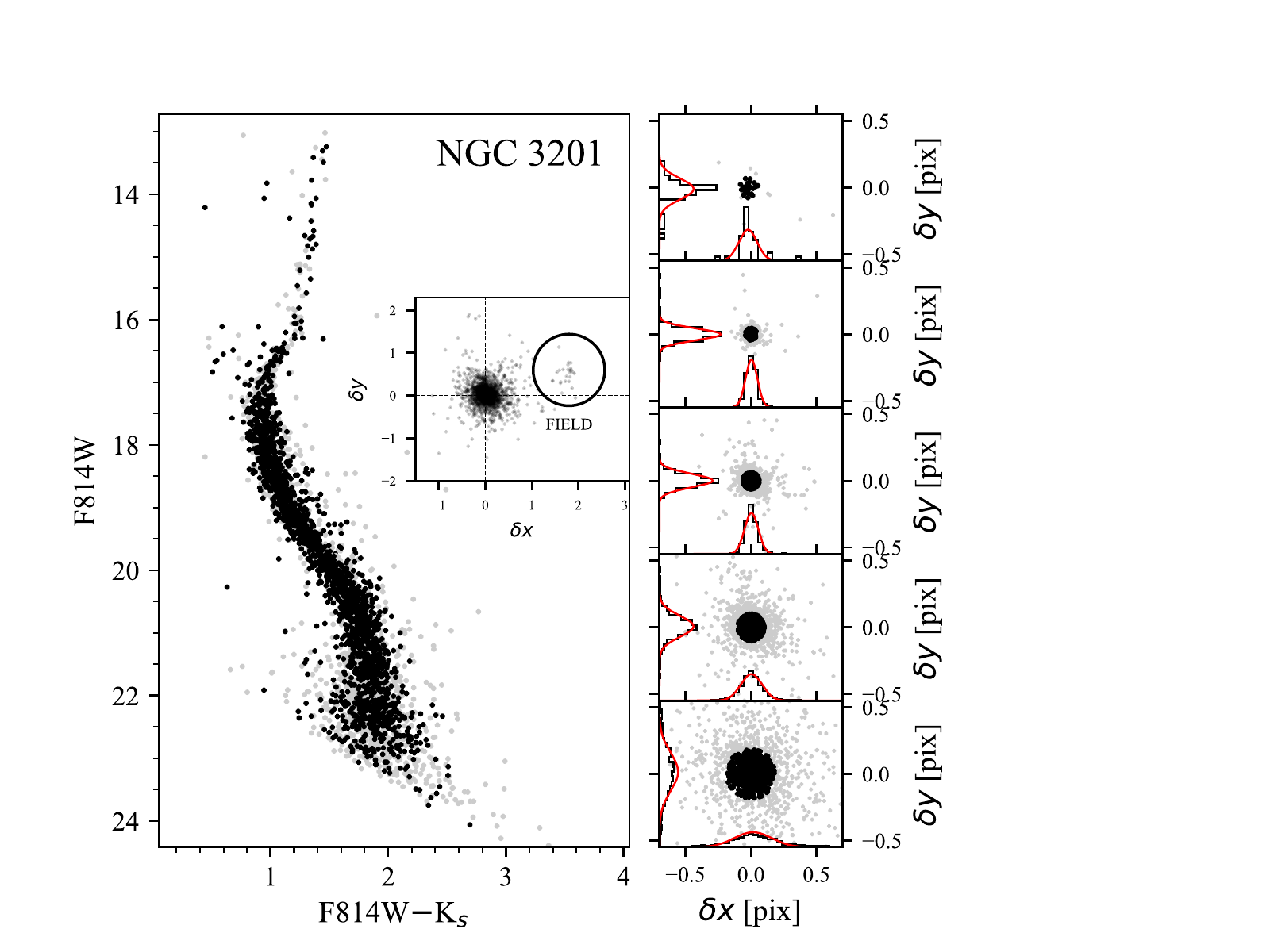}
   \includegraphics[width=0.49\linewidth]{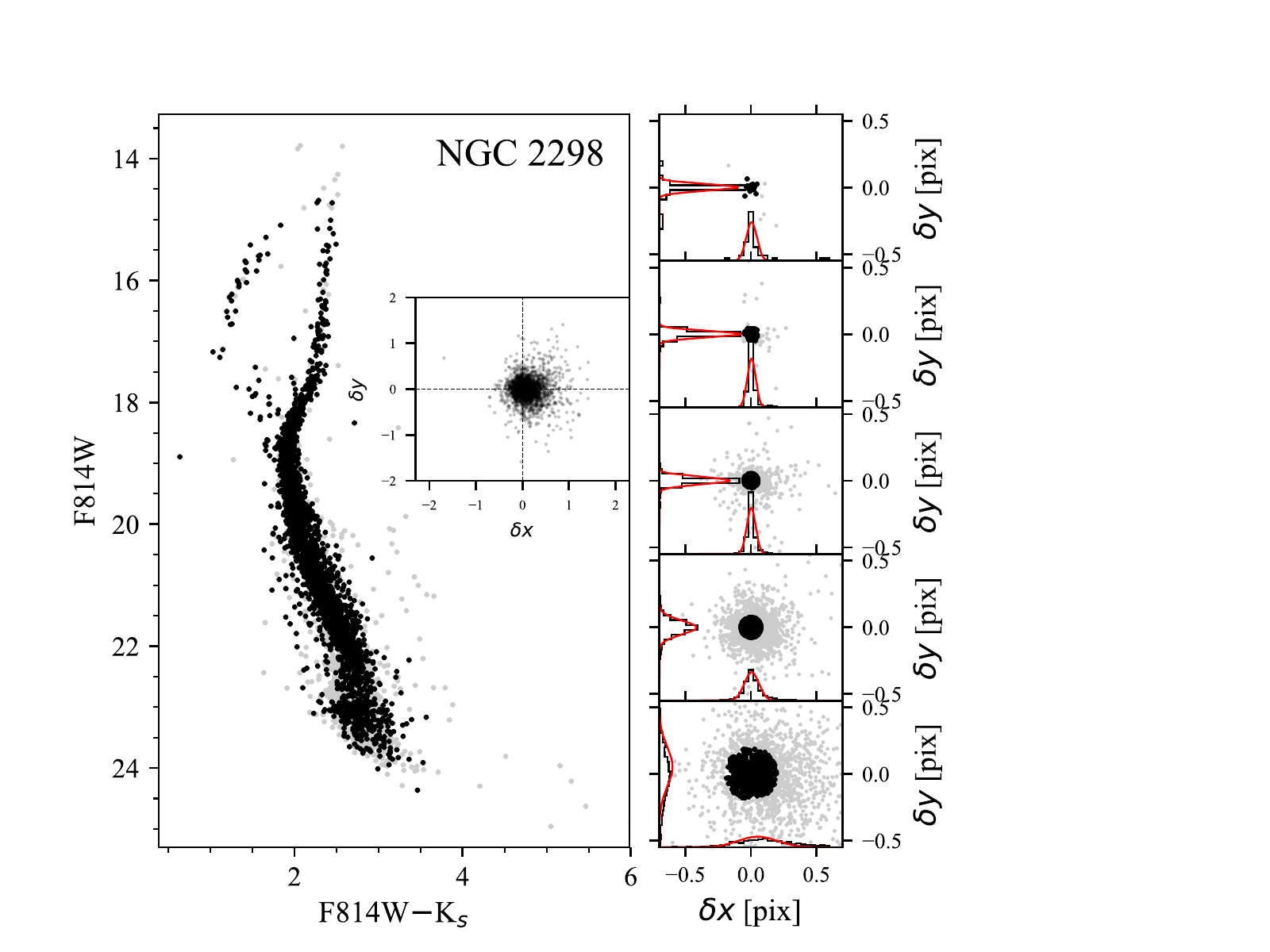}
   \caption{HST-Gemini CMDs of cluster members (black points) and non-members (grey points; also shown in Figure~\ref{fig:pm_error}) for NGC\,3201 ({\it left}) and NGC\,2298 ({\it right}), as measured in the combined optical-NIR filters.~The vector point diagrams are shown for five magnitude bins in each cluster, with the corresponding color coding for selected cluster members and non-members stars.~The black lines in each sub-panel show the density histograms for horizontally and vertically binned data.~The red lines show Gaussian probability density function fits to the binned data.~Additional inset panels in the CMD plots show the vector point diagrams of the whole catalog for each cluster.}
   \label{fig:pm_selection}
\end{figure*}

The precision of PM measurements is largely determined by the ability to remove the different distortion effects in the camera optics.~Unfortunately, the different distortions present in MCAO instruments are very hard to model from first principles given their complexity, since they likely depend on the specific observing conditions and instrument configurations (e.g.~offsets, relative position of the natural and laser guide star constellation, etc.).~For the case of GeMS/GSAOI, in addition to anisoplanatism, other distortions caused by gravity flexure or movement of the AO-bench have been suggested \citep{nei14b}.~In a recent study by \cite{mas16}, combined GeMS/GSAOI and HST/ACS data were used to measure the astrometric capabilities of the GeMS/GSAOI instrument.~The authors used previous HST/ACS PM measurements in the NGC\,6681 field to isolate and directly measure the distortion effects in GeMS, which they report to reach maximum amplitudes between $-1.64$ and 4.83 pixels in $x$ and between $-0.49$ and 0.35 pixels in $y$.~Given these considerations, we chose to remove such distortions using a chip-by-chip and exposure-by-exposure approach.~However, we point out that at no point were we interested in or attempting to measure absolute PMs.~Our analysis aims at reducing the impact of GSAOI field distortions in order to obtain relative PMs for the decontamination of cluster member stars, as well as testing the astrometric capabilities obtained from the experimental design of our survey.

To begin with, we cross-matched the stars in both epoch catalogs using the approximate RA and Dec coordinates obtained from the GSAOI image header.~In the case of NGC\,3201, the transformation was calculated using the stars classified as cluster members in \cite{sim16}.~For NGC\,2298 we used the preliminary catalog of the HST UV Legacy Survey of GCs \citep{sot17, pio15} to select stars with displacement values $D_x, D_y\!=\!0$ \citep[see Table 2 in][]{sot17}, i.e.~stars whose PMs are consistent with the cluster's motion.~This way we effectively removed large intrinsic stellar motions, leaving only the cluster's mean motion and the distortion effects left to be fit.~We then used the {\sc geomap} task from {\sc IRAF} to calculate geometric transformations in order to map the GSAOI ($x,y$) pixel coordinate system onto the reference ACS ($x,y$) pixel coordinate system.~Within the {\sc geomap} task, we used ``general" for the fit geometry.~With this option, a linear term and a distortion term are computed separately.~The linear term includes a $x$ and $y$ shift, a $x$ and $y$ scale factor, a rotation and a skew parameter.~The distortion term consists of a third-order polynomial fit to the residuals of the linear term, therefore increasing the number of coefficients to 9 per axis.~We also included the option for a 3$\sigma$-clipping iteration on the residual values in order to remove any remaining outliers.~The transformation was then applied with the {\sc geoxytran} task within {\sc IRAF} to the entire GSAOI catalog, chip-by-chip and exposure-by-exposure.~Based on previous works which have studied the geometric distortions of the GeMS/GSAOI field \citep{nei14b,mas16,dalessandro16} we expected our transformed coordinate system to be distortion-corrected down to $\sim$1 mas. In particular, \cite{nei14b} found that for single-epoch, non-dithered data, astrometric precision of $\sim$0.2 mas was possible for bright stars and 60 sec exposures.~For our case of dithered data with 30 sec exposures, they predict around $\sim$0.2$-$0.5 mas precision (see their Figure 17), while both \cite{mas16} and \cite{dalessandro16} report an astrometric accuracy of $\sim$1 mas for high stellar density fields.~Therefore, we expected to reach a precision in the $\sim$0.2$-$1.0 mas range which we considered to be sufficient for our intended relative PM decontamination.

To explain our method of PM cleaning, let us define ($x'_{ij},y'_{ij}$) as the transformed GSAOI star coordinates for chip $i$ and exposure $j$, and ($x_{\rm ref},y_{\rm ref}$) as the reference ACS star coordinates.~We combined all exposures for chip $i$ and calculated the average positions, ($\overline{x'}_{i},\overline{y'}_{i}$).~The error of the average value for a given coordinate $\nu$ was then calculated as $\sigma_{\overline{\nu}}=\sqrt{\sigma_s^2/N}$, where $\sigma^2_s=(N\!-\!1)^{-1}\sum_{k}^{N}(\nu'_k-\overline{\nu'})^2$ is the unbiased estimate of the sample variance and $N$ is the number of detections.~The final catalog was then a combination of all chips, selecting only stars with at least 4 detections, which provided the coordinates ($\overline{x'},\overline{y'}$).~In Figure~\ref{fig:pm_error} we show the obtained astrometric error\footnote{Pixel units were converted to mas units by using the plate scale 0.05\arcsec/pix that corresponds to the ACS/WFC camera.} for stars in NGC\,2298 and NGC\,3201 as a function of magnitude (grey and black points are explained further below).~In both cases we observed an astrometric precision consistent with the expected performance as explained earlier, although the degradation of the PSF in the NGC\,3201 data (see Section~\ref{sec:lgswfsm}) evidently contributed to comparatively worse astrometric precision.~However, it is clear that with good AO correction one can achieve sub-mas astrometric precision down to F814W~$\simeq20$ mag within a few minutes on GSAOI.

We then calculated the PMs in ACS pixel units $\delta x=\overline{x'}-x_{\rm ref}$ and $\delta y=\overline{y'}-y_{\rm ref}$ and used them to construct vector point diagrams.~We did this by separating the entire luminosity range into 5 magnitude bins and constructing vector point diagrams for each bin.~The cluster members selection was then done as follows; first, we calculated the standard deviation of the value $R\!=\!\sqrt{\delta x^2+\delta y^2}$ for each bin, then selected as cluster members only those stars that had $R\!<\!1\sigma_R$, for each individual magnitude bin.~This conservative threshold helped us to reject the stars with a comparatively worse PSF and astrometric precision, such as bright saturated stars and stars near the edge of the detectors.

The results of the PM selection are illustrated in Figure~\ref{fig:pm_selection} for NGC\,3201 and NGC\,2298, where the black and grey points show the selected members and non-members, respectively, as also shown in Figure~\ref{fig:pm_error}.~The vector point diagrams for each magnitude bin are shown accordingly in the sub-panels next to each CMD.~The black lines in each sub-panel show the density histograms for horizontally and vertically binned data.~The red lines show Gaussian probability density function (PDF) fits to the binned data.~Indeed, the measured PMs of the central clump in the vector point diagrams are typically well fit with a Gaussian PDF, which supports our cluster members selection using a sample dispersion estimate such as the standard deviation.~Note that the distribution becomes broader for fainter (i.e. less massive) stars\footnote{The brightest bin can be ignored since it has a relatively low number of stars, in addition to the star's measured centroids being more affected by saturation.}, which is expected both from photometric errors and energy equipartition.~This justifies the use of separate vector point diagrams at different magnitude bins for the cluster members selection.~Also, we note that the CMDs are almost unaffected by the PM cleaning, which is explained by the seemingly low contamination of field stars inside the fields covered by the GSAOI data.~Only for NGC\,3201 we are able to clearly distinguish a small number ($\sim$30) of field stars\footnote{This is due to the lower galactic latitude of NGC\,3201, i.e. $b=8.64^{\rm o}$, compared to $b=-16^{\rm o}$ for NGC\,2298.} in the data, as shown in the inset panels.~Moreover, the rejected stars (grey points) in Figure~\ref{fig:pm_selection} are shown in Figure~\ref{fig:pm_error} to discriminate what is our rejection based on.~In the case of NGC\,2298, the relatively superior astrometry means that we were able to recover the full intrinsic stellar motion distribution for the brighter bins and reject mostly the few objects with a poorer astrometric precision. In contrast, for the fainter magnitudes the PM threshold becomes more strict than the full stellar motion distribution and hence we started rejecting at the tails of the velocity distribution. In the case of NGC\,3201, the relatively poorer astrometry and closer heliocentric distance (see Table~\ref{tab:gcprop}) results in our PM threshold always being comparable (or smaller) to the full intrinsic stellar motion distribution and hence the selected members and non-members share the same distribution in Figure~\ref{fig:pm_error}, as one would expect for a non-biased selection.~In conclusion, a large fraction of the rejected stars likely corresponded to actual cluster members on the tails of the velocity distribution and, hence, these particular cluster-member samples are not representative of the full internal dynamical distribution, but are rather more robust non-contaminated samples.~The main goal of including the PM measurements in this paper is for the opportunity to test the astrometric capabilities obtained from the experimental design of our survey, as demonstrated in the next paragraph. 

\begin{figure}[t]
   \centering
   \includegraphics[width=\linewidth]{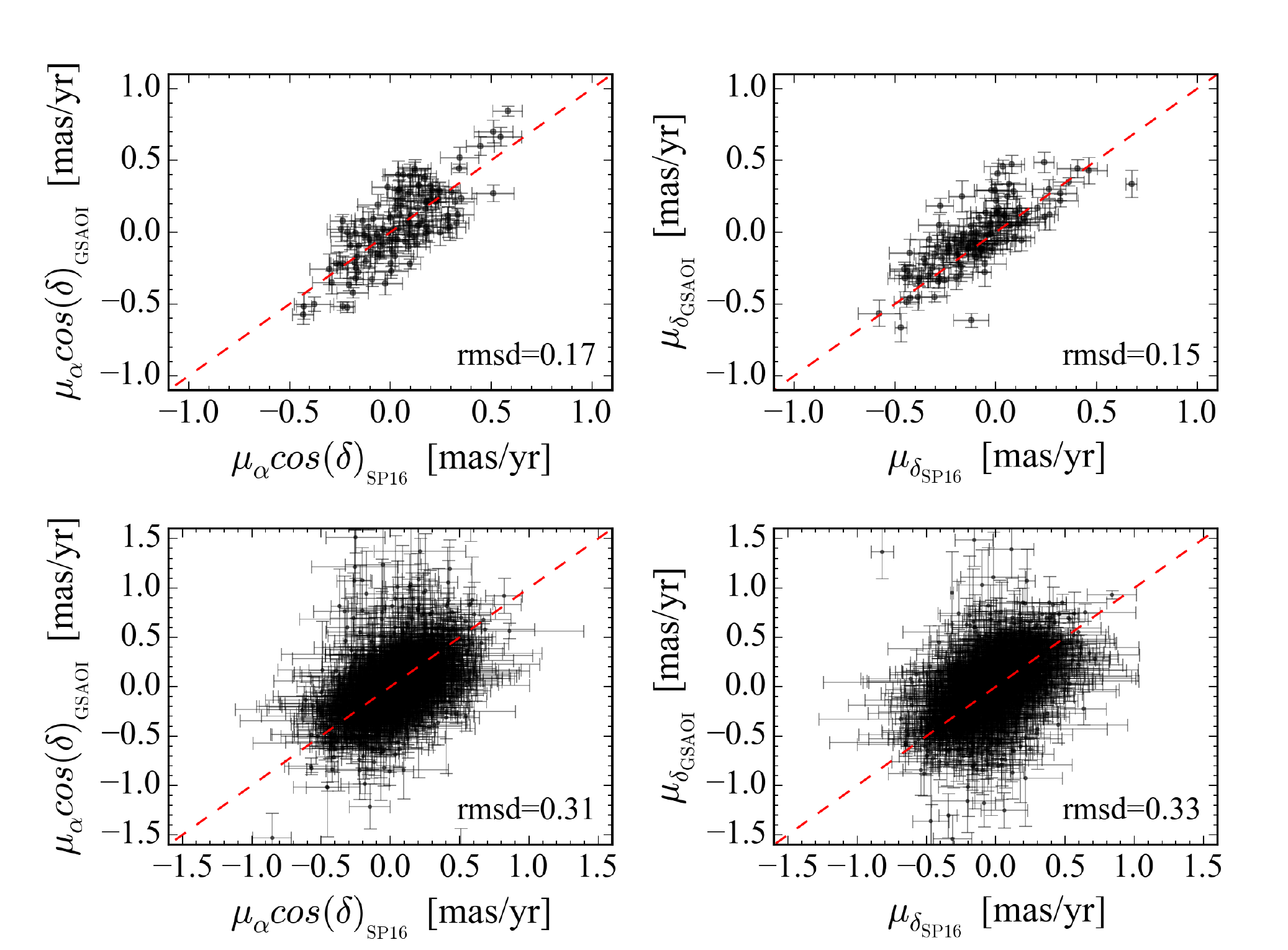}
   \caption{PMs measured in sky coordinates units in NGC\,3201 versus the ones measured by \cite{sim16} for a sample of stars in common.~The two top panels show only stars with PM errors $\sigma\leqslant$ 0.1 mas/yr.~The observed scatter around the linear relations are shown in the panels as the root mean squared deviation values.~The scatter increases by a factor of two for the lower panels, which include all stars in common.~The panels show good agreement between both studies, and the scatter observed is consistent with the estimated errors.}
   \label{fig:pm_comparison}
\end{figure}

In a recent work, \cite{sim16} measured PMs using multi-epoch HST imaging data for the central regions of a large sample of GCs.~Their sample included NGC\,3201, which allowed us to compare the PM measurements for all stars in common in both studies.~We transformed all PMs from the ACS/WFC pixel units into the more common milli-arcseconds (mas)/yr units by assuming the ACS/WFC pixel scale equal to 0.05\arcsec/pix and divided by the time baseline ($\sim\!7$ years) between the first and second epoch.~We then compared the PMs measured for each sky coordinate in Figure~\ref{fig:pm_comparison}.~The two top panels show only a sample of stars with relatively well constrained PMs.~Specifically, we include only the stars with PM errors $\sigma\leqslant$ 0.1 mas/yr, as calculated using the formula previously shown.~Adding in quadrature the maximum error contribution from both measurements gives $\small{\sqrt{(0.1)^2+(0.1)^2}}=0.14$ mas/yr, which agrees nicely with the measured scatter around the linear relation, shown in the panels as the root mean squared deviation values.~The scatter increases by a factor of two for the lower panels, which includes all stars in common.~A more detailed analysis on this comparison is beyond the scope of this work, but this simple test shows that the measured PMs in this work can reach a similar accuracy to the ones in \cite{sim16}, which are entirely based on HST data.~This suggests a promising potential of the astrometric capabilities of the GeMS/GSAOI instrument when combined with HST-based data, as found also by recent similar studies \citep{mas16, dalessandro16}.

The PM-cleaned catalogs of NGC\,3201 and NGC\,2298 are used throughout the rest of this work for the characterization of the CMDs and comparison with stellar evolutionary models.

\subsection{Differential Reddening}\label{sec:dif_red}

\begin{figure}[t]
   \centering
   \includegraphics[width=\linewidth]{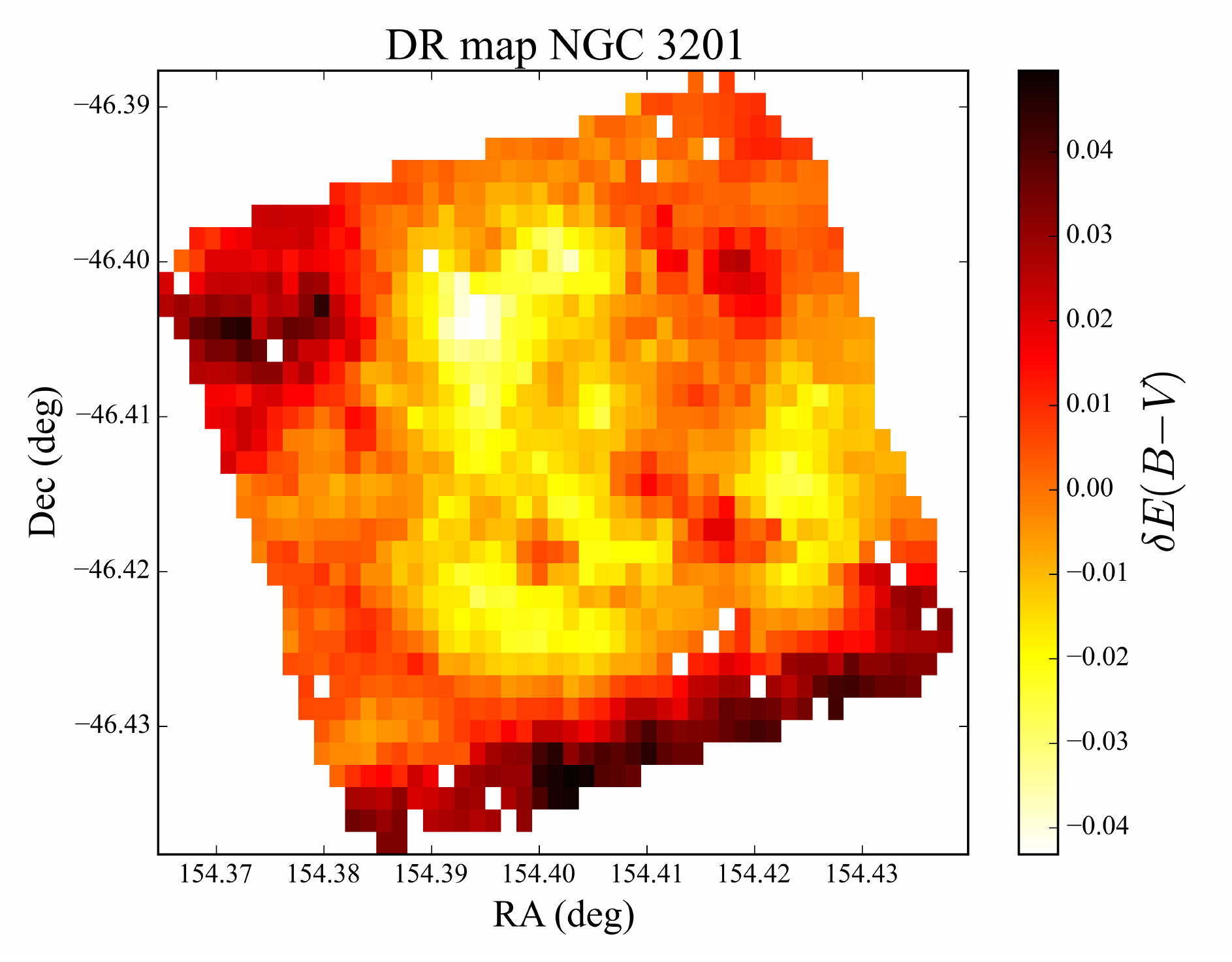}
   \includegraphics[width=\linewidth]{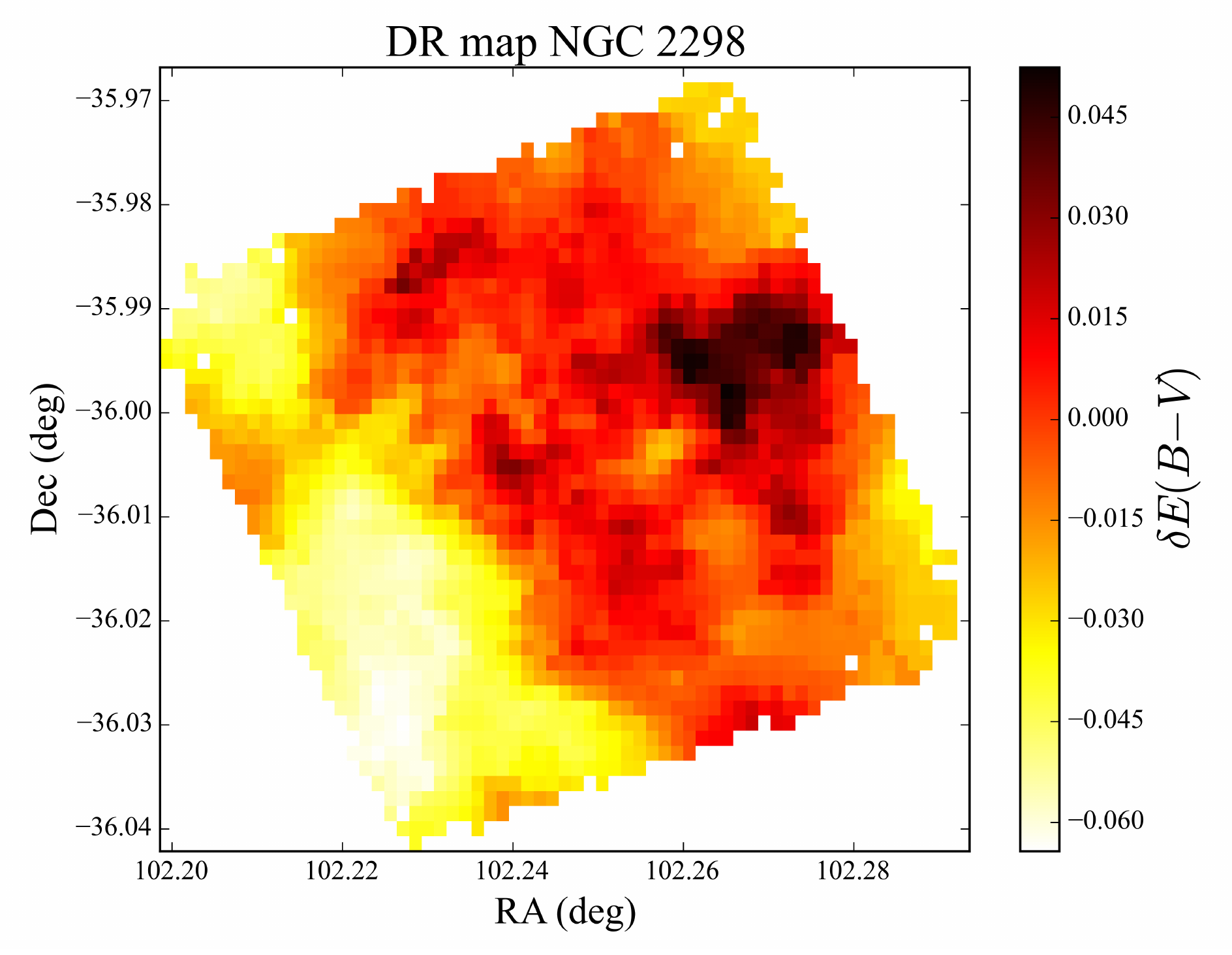}
   \caption{Differential reddening (DR) maps for NGC\,3201 ({\it top}) and NGC\,2298 ({\it bottom}). Both star catalogs were divided into bins of 0.09\arcmin\ in both sky coordinate axes.~The color map represents the average value of $\delta E_{B-V}$ for a given binned region of the field.}
   \label{fig:DRmaps}
\end{figure}

\begin{figure*}[t]
   \centering
   \includegraphics[width=0.495\linewidth]{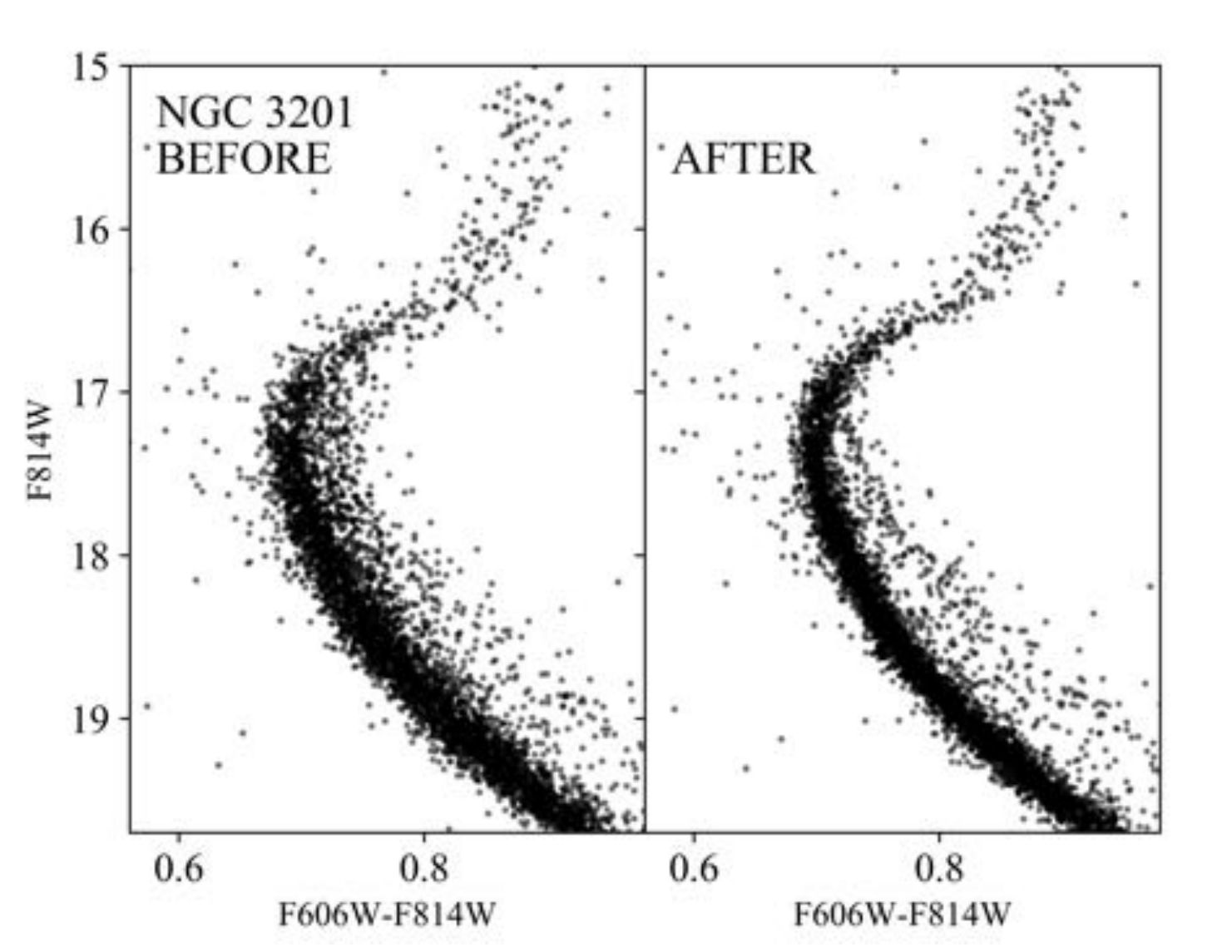}
   \includegraphics[width=0.495\linewidth]{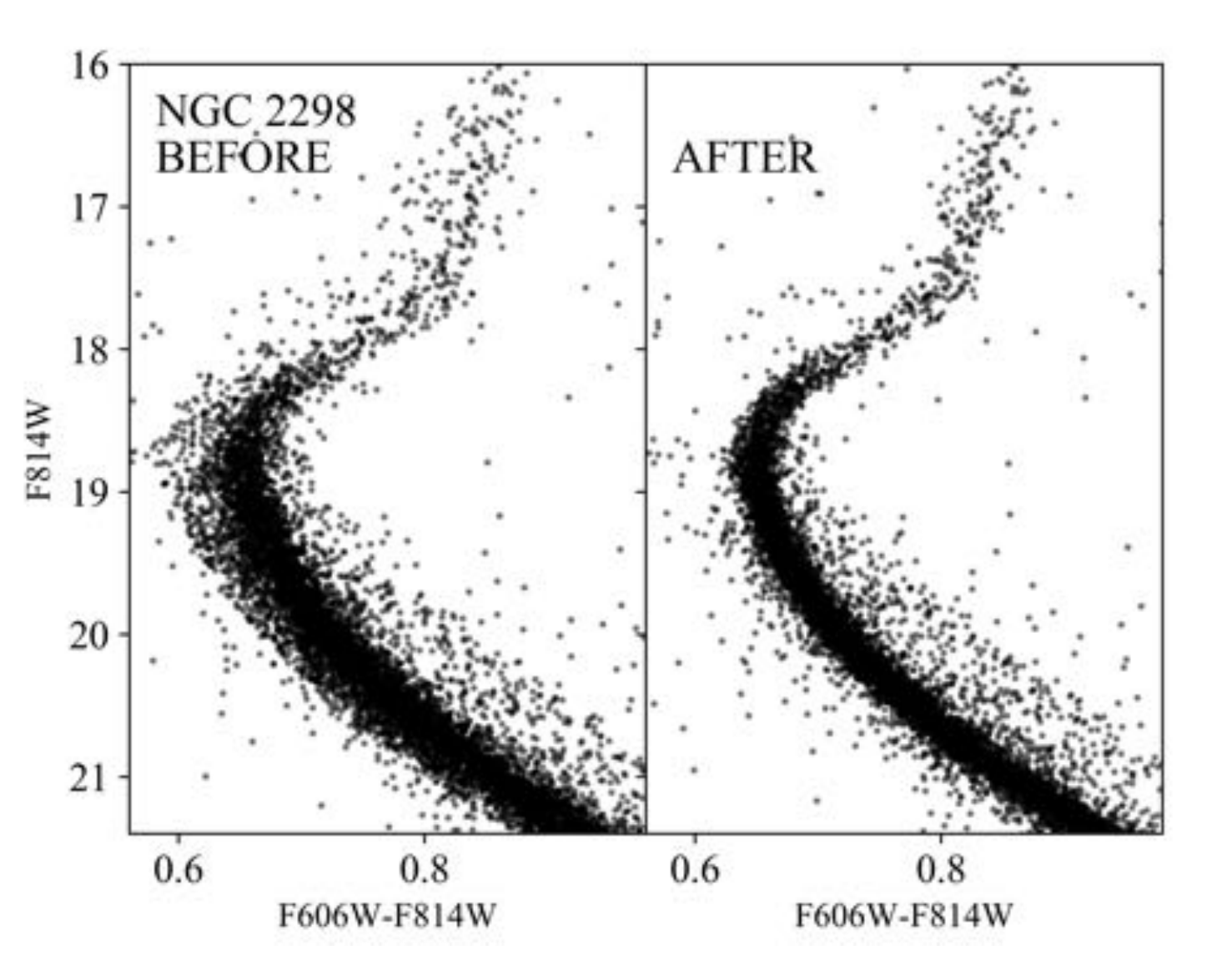}
   \caption{\textit{Left}: Two panels showing the CMD of NGC\,3201 before and after the DR correction.~\textit{Right}: Same as left panel for NGC 2298.~The CMDs are zoomed to show the region around the SGB and upper-MS, where the features are more notably enhanced after the DR correction.}
   \label{fig:before_after_DRC}
\end{figure*}

Prior to isochrone-fitting and determination of the best-fit isochrone, we determined the differential reddening (DR) corrections for all optical (F606W and F814W) and UV (F336W and F438W) bands.~No DR correction was applied to the NIR filters, as this was found to be smaller than or comparable to the photometric uncertainties for those filter magnitudes.~The choice to determine DR corrections was made in an effort to minimize the contribution of external errors to any detected spread in color-space.~By accounting and correcting for a spread in color-space due to any DR, the accuracy and fidelity of subsequent isochrone-fitting was improved.~Due to the need for an abundant number of stars and a well defined fiducial line in the CMD, we calculated the effects of DR using the \cite{sot17} HST catalogs for both clusters, utilizing the larger HST FoV.~To determine the DR corrections applied to our catalogues, we followed a method inspired by previous literature.

We chose to adopt the methodology presented in \cite{mil12}.~This method consists of measuring the local DR affecting one single cluster star, by measuring the median shift along the reddening vector in a sample of nearby MS stars, then repeating this measurement for every single star in the cluster catalog.~The median shift is measured with respect to the MS fiducial line, using the 20 nearest MS stars to the target star for which we calculated the local DR correction.~As described in \cite{mil12}, the median shift along the reddening vector is best seen when the CMD axes were rotated to align the horizontal axis with the reddening vector.~The axes were rotated counterclockwise around the point that corresponded to the MSTO, by an angle defined as:

\begin{equation}
\theta = \arctan\left(\frac{A_{\rm F814W}}{A_{\rm F606W}-A_{\rm F814W}}\right)
\end{equation}

\noindent where the extinction coefficients $A_{\rm F606W}$ and $A_{\rm F814W}$ are obtained by interpolating the extinction values for HST/ACS filters from Table 3 in \cite{bedin05}.~The interpolated extinction values are calculated for the corresponding $E_{B-V}$ of NGC\,3201 and NGC\,2298 found in the McMaster (Harris) Catalog \citep{har10}.~Once the CMD axes are rotated, a measurement of the shift in values with respect to the fiducial line is taken directly from the x-axis.~The median shift value is then calculated for the sample of nearby MS stars and taken as the local DR, which is then subtracted from the target star.~An illustration of the previous steps is shown in Figure 6 of \cite{mil12}.

By applying this process to all the stars, we generated a more defined MS before calculating the fiducial line in the rotated CMD again.~The procedure was iterated through two more additional times, after which we assumed a fully DR-corrected CMD.~The coordinates were then transformed back to the original F814W vs.~F606W--F814W plane.~By comparing the corrected magnitudes to the non-corrected magnitudes, we used the extinction-law cited above to calculate the DR $\delta E_{B-V}$ for each star and to construct a DR map for our two cluster fields.~The DR maps are shown in Figure~\ref{fig:DRmaps}.~Using the $\delta E_{B-V}$ values and the extinction law from Table 3 in \cite{bedin05} and \citet[for $J, H$ and $K_s$]{case14}, we calculated the DR corrected magnitudes for each star, for the remaining filters: F336W and F438W.~As a summary of the described process, we show the CMDs of NGC\,3201 and NGC\,2298 before and after DR correction in Figure~\ref{fig:before_after_DRC}, illustrating the ability of the method to effectively accentuate the features of the CMD.
 
\subsection{Photometric Catalogs}\label{sec:syn_cat}
Following the calibration of the near-IR photometry PM-cleaning and DR-correction, master catalogs were created combining the three GSAOI near-IR bands $JHK_s$ together with the two HST/ACS optical filters F606W(V) and F814W(I) from \cite{sar07}, as well as the intermediate-data release photometry in the near-UV filter F275W and the two optical filters F336W(U) and F438W(B) from the HST/WFC3 \textit{UV Legacy Survey} \citep{pio15}.~Mean reddening corrections were not applied directly to the catalogs.~Instead, the mean reddening values were determined during the isochrone-fitting procedure described below.~During this process we assumed extinction ratios $R_{i}$ given by \cite{case14}, which yielded reddening-corrected stellar magnitudes via $m_{i,0}=m_{i}-R_{i} E_{B-V}$, where $m_i$ is the measured magnitude and $E_{B-V}$ the corresponding reddening.~Table~\ref{tab:ebv} summarizes the $R_{i}$ values for each filter.~Caveats associated with this choice will be discussed in subsequent sections.

\begin{deluxetable}{lccccccc}
\tabletypesize{\footnotesize}
\tablecaption{Extinction ratios from \citep{case14}. \label{tab:ebv}}
\tablehead{
\colhead{} & \colhead{F336W} & \colhead{F438W} & \colhead{F606W} & \colhead{F814W} & \colhead{$J$}& \colhead{$H$}& \colhead{$K_s$}
}
\startdata
$R_{i}$ & 5.148 & 4.135 & 2.876 & 1.884 & 0.899 & 0.567 & 0.366 \\   
\enddata
\tablecomments{The reddening corrections are defined as: $m_{i,0}=m_{i}-R_{i}E_{B-V}$, where $E_{B-V}$ is determined in the following sections and the extinction ratios $R_{i}$ for each filter $i$ are the tabulated values.}
\end{deluxetable}

%%%%%%%%%%%%%%%%%%%%%%%%%%%%%%%%%%%%%%%%
\section{Results} \label{sec:results}
In the following section we utilized the master photometry catalogues to perform absolute age determinations, derive distances, reddening values, and chemical characteristics for the two target clusters.~Combining the GSAOI and HST photometry proved to be the most diagnostic combination for determining the cluster ages.~This is attributed mainly to the clear appearance of the MSK in the faintest measured magnitude regime.~As mentioned previously, the combination of the MSK and MSTO is independent of cluster distance and reddening and can, therefore, be used for absolute age determination.~We note that earlier studies based on AO-supported imagers, such as VLT/MAD \citep{mar07}, LBT/PICES \citep{esp10}, and Gemini/GSAOI have used similar purely ground-based near-IR photometry \citep{sara16} or combinations of ground- and space-based photometry to study MW GCs \citep[e.g.][]{mor09, bon10, tur15, mon15, sara15, mas16}.~However, these studies were either not deep enough to sample the MSTO {\it and} MSK simultaneously with high-enough photometric quality or did not perform PM-cleaning.~Our study combines all of these advantages, for the first time.

\subsection{Isochrone Models and Stellar Population Parameters}\label{sec:chi}
We explore the model dependencies in the age determinations by using the following isochrone sets:
\begin{itemize}

\item Dartmouth Stellar Evolution Database \citep[\texttt{DSED};][]{dot08} with [Fe/H]~$=(-1.59, -1.92)$ dex, $Y=0.245$ and [$\alpha$/Fe]~$=0.2$ dex.

\item Victoria-Regina Isochrone Database \citep[\texttt{VR};][]{vdb14} with [Fe/H]~$=(-1.59, -1.92)$ dex, $Y=0.25$ and [$\alpha$/Fe]~$=0.2$ dex.

\item A Bag of Stellar Tracks and Isochrones \citep[\texttt{BaSTI};][]{piet04} with [Fe/H]~$=(-1.60, -1.90)$ dex, $Y=0.246$ and [$\alpha$/Fe]~$=0.2$ dex (generated specifically for this work.)

\end{itemize}
We initially used all three isochrone sets to find the CMD regions which showed the maximum dynamic range (i.e.~spread in magnitude and color relative to their corresponding uncertainties) prior to determining the age of the GCs.~From this study we found that the best filter combination to perform absolute age determination was the $K_{s}$ vs.~F606W$-K_{s}$ combination, a result of the photometric quality of the data in these bands and the ability to \textit{narrowly} detect the MSK.~We also chose to use the F336W vs.~F336W$-K_s$ combination due to its added diagnostic power, particularly in the sub-giant branch region.~In the following section, we use these filter combinations to determine the absolute ages of both target GCs, through an exploration of the stellar population parameter space and subsequent pseudo-$\chi^{2}$ minimization.

\begin{figure*}[t]
   \centering
   \includegraphics[width=\linewidth]{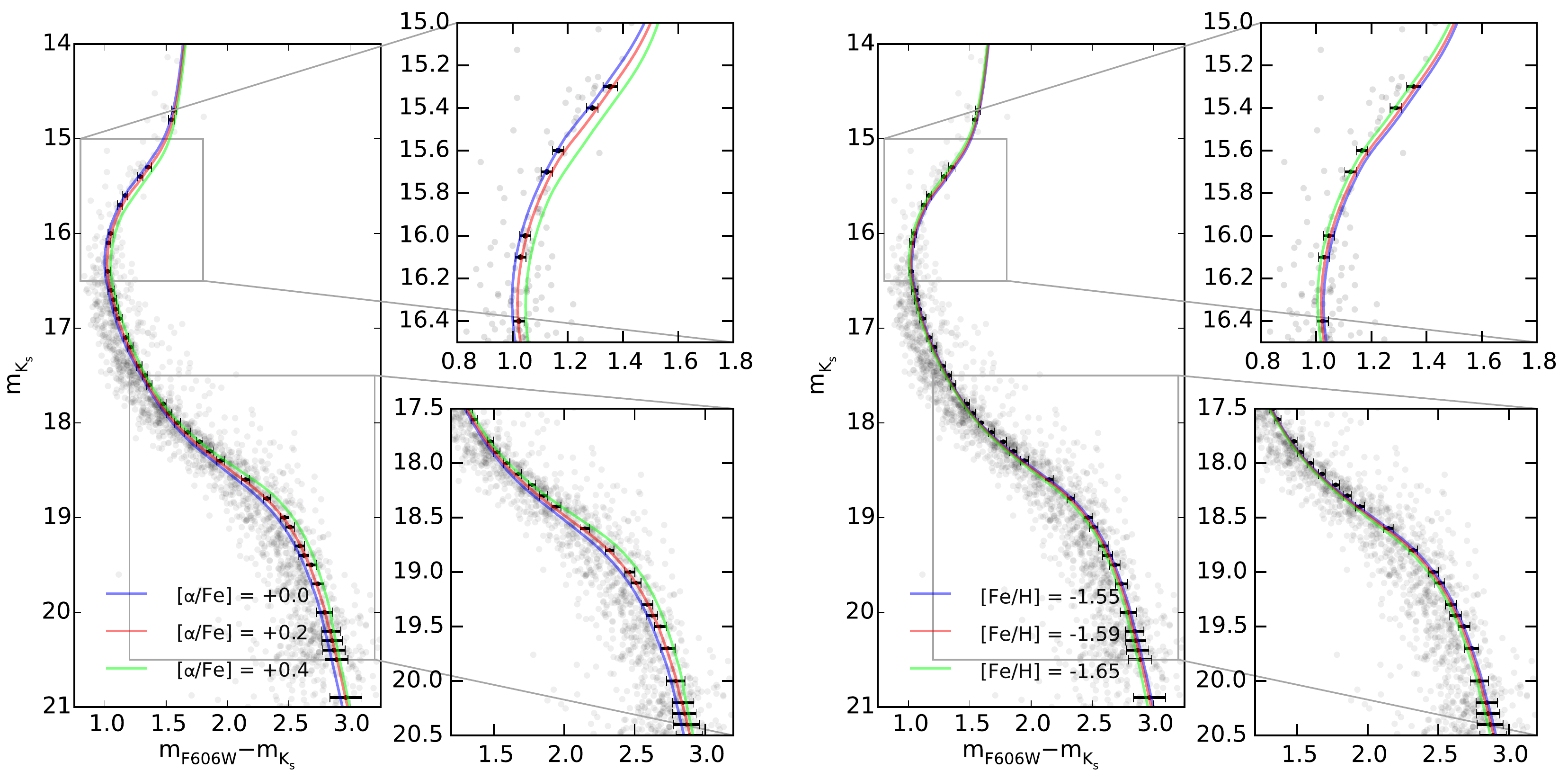}
   \caption{Illustration of the fit dependencies on $\alpha$-element enrichment ({\it left panel}) and metallicity ({\it right panel}) with focus on the SGB and MSK regions, for the case of NGC\,3201.~Overlaid isochrones were created using the \texttt{DSED} isochrone sample for a best-fit age of 12.2 Gyr, distance of 5.1 kpc, and $E_{B-V}=0.25$\,mag (see Section~\ref{sec:chi}).~Values of metallicities and [$\alpha$/Fe] are listed in the plots, with error bars representing the photometric error associated with a random data point at the given magnitude near the best-fit isochrone.}
   \label{fig:iso_dep}
\end{figure*}

We chose to fix the cluster metallicities and \afe\ enrichments using the McMaster Catalog \citep{har10} and \cite{dot10} respectively, following preliminary isochrone fitting to visually assess the fit to the data around the literature values.~From this initial investigation, we found that fits to the MS depended weakly on metallicity, while fits to the MSK were somewhat dependent on \afe\ but the chosen value of $[\alpha/\mathrm{Fe}]\!=\!0.2$ demonstrated a good match to the data.~These dependencies are shown for the case of NGC\,3201 in Figure~\ref{fig:iso_dep}.~We do, however, acknowledge the introduction of possible systematic errors related primarily to the choice of an inaccurate $\alpha$-element enrichment of the isochrone set.~In future papers, a full exploration of both chemical and physical parameters will be done for a larger sample of clusters in an effort to limit the effects of any \textit{a-priori} assumptions.

In the following, we explore only the parameter space of distance, reddening, and age to determine the best values and uncertainties in each parameter. The ranges of possible distances and values of reddening were centered about the values given in Table~\ref{tab:gcprop}.~The parameter space exploration was performed in steps of 0.2 kpc in distance and 0.1 mag steps in reddening.~Additionally, initial guesses for the ages of $12^{+2}_{-2}$ Gyr and $13^{-1}_{+2}$~Gyr were chosen for NGC\,3201 and NGC\,2298, respectively, based on recent age determinations \citep{dot10, mun13, wag16}.

\begin{figure*}[t]
   \centering
   \includegraphics[width=\linewidth]{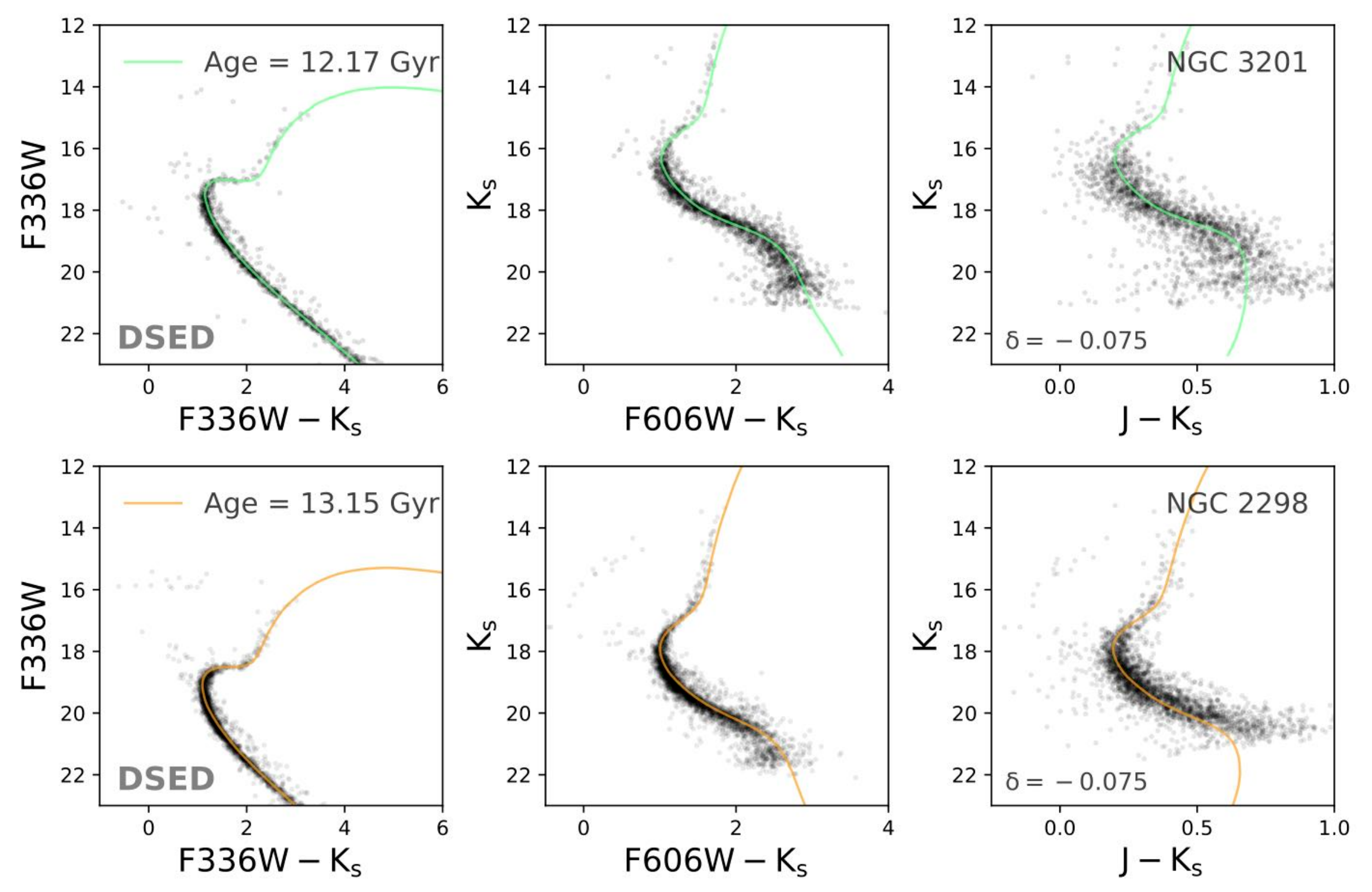}
   \caption{Illustration of the MSK-MSTO-SGB morphologies in various CMDs for NGC\,3201 ({\it top panels}) and NGC\,2298 ({\it bottom panels}).~We used the best-fit isochrone from the \texttt{DSED} library for both clusters using the derived values of reddening, distance, and stellar population parameters shown in Table \ref{tab:chi2} based on our PM-cleaned and DR-corrected catalogues (see text for details).~We point out that an offset of $\delta(J\!-\!K_s)\!=\!-0.075$\,mag has been applied to the pure near-IR isochrones for both clusters in order to better match the isochrone predictions with the MS.}
   \label{fig:all_cmds}
\end{figure*}

\subsection{A Note on the Use and Nomenclature of the MSK}\label{sec:MSKnote}
Throughout this paper, we reference the appearance and recovery of the MSK in the optical-near-IR color combinations of our two clusters.~Although we do observe the MSK in both clusters in the optical-near-IR color combinations, we do acknowledge that for the purpose of isochrone fitting we do not recover enough of the knee to anchor our isochrones using this feature alone.~Instead, we make use of the so-called ``MS-saddle", the point of minimum curvature on the lower MS, to help us anchor our isochrones in the low luminosity regime.~As discussed in \cite{sara18}, the MS-saddle point is often mis-identified in the literature as being coincident with the MSK.~As \citeauthor{sara18}~explain, the true MSK can only appear completely in pure near-IR CMDs and is a fixed physical feature, while the MS-saddle is a purely geometric feature.~The authors also discuss the limitations of current isochrone models to accurately predict the position of the MSK, showing that the \texttt{DSED}, \texttt{VR} and \texttt{BaSTI} models predict different locations of the MSK in the same color combinations.~They attribute this disagreement to differing model boundary conditions, leading to different effective temperatures in the regime of the MSK.~However, they also find that the three models predict the same location for the MS-saddle, but that each model is sensitive to a different parameter influencing the final age determination.

As an example, the \texttt{BaSTI} isochrones were found to be most sensitive to changes in distance between the MSTO and MS-saddle, while the \texttt{DSED} isochrones were found to be most sensitive to changes in metallicity \citep{sara18}.~As we were limited to 0.5 dex sampling in the creation of our \texttt{BaSTI} isochrones, but not our \texttt{DSED} isochrones, we consider this to be a positive finding.~In the following sections we discuss our methodology for fitting the entire MS as part of our age determination, extending down to the what we consider to be the recovered MSK.~We do, however, point out that a weighting system is introduced to exploit the better modeled MS-saddle and MSTO.~As a means of error-minimization, we create a ridgeline using overlapping windows, define the MS-saddle using the analytical definition alone, as opposed to a geometric method \citep[see][]{mas16}, and take into account the photometric errors associated with each data point during final pseudo-$\chi^{2}$ isochrone fitting.~Finally, we caution that the accuracy of following age determinations are subject to the accuracy of both the MS ridgeline and the models themselves.

\subsection{Methodology of Isochrone Fitting}\label{txt:isofit}
To refine the range of possible chemical characteristics, and to derive distance and reddening values for the GCs, a study similar to that of \cite{sara16} was performed.~We optimized the isochrone fits associated with fixed values of [Fe/H] and \afe\ in the aforementioned age ranges by allowing the values of reddening and distance to wander away from the literature values until a true minimum in color-magnitude space was found.~In this way we chose to minimize the \textit{distance} in color-magnitude space between each isochrone set and the fiducial isochrone constructed from the data.

\subsubsection{Empirical Isochrone Ridgeline Construction}\label{sec:ridge}
Prior to the color-magnitude distance minimization, a SGB-MS ridgeline, or fiducial isochrone, was constructed, representative of single stars with the most accurate photometry.~This was done in order to remove the contamination in color space attributed to stars with larger photometric errors and the binary population.~To create the fiducial isochrone a sliding window was generated in magnitude space, with the window size dependent on the filter combination used for each particular CMD.~Seventy stars were found to give robust intra-window statistics, while allowing for efficient rejection of outliers in the $K_s$ vs.~F606W$-K_s$ color combination, whereas forty was the corresponding number for the F336W vs.~F336W$-K_s$ CMD.~For each window position, the median magnitude, color and associated uncertainty given by the interquartile range (IQR) of the star sample was recorded as part of the fiducial isochrone creation.~After recording said values, the sliding window progressed downwards by one star towards fainter magnitudes.~This procedure was then repeated sliding along the SGB and MS for different desired color combinations, for both clusters.~The resulting ridgelines were then smoothed using kernel sizes of 0.01 magnitudes and examined visually in all desired color combinations to ensure the fidelity of this fiducial isochrone modeling technique, specifically in the presence of significant binary sequences (see Fig.~\ref{fig:all_cmds}).

\begin{deluxetable*}{ccccccc}
\tablewidth{0.75\textwidth}
\tablecaption{Locations of the most diagnostic regimes (in apparent magnitudes) in the two chosen color-magnitude combinations and the chosen weighting function coefficient.\label{tab:msto_loc}}
\tablehead{
\colhead{Cluster}   &\colhead{CMD}  &\colhead{$m_{\text{MSTO}}$}    &\colhead{$m_{\text{MS-saddle}}$} &\colhead{$\Delta(m_{\text{MSTO}}- m_{\text{MSK}})$}    &\colhead{$m_{\text{SGB}}$} &\colhead{$a$}}
\startdata
NGC\,3201   &$K_{s}$, $F606W-K_{s}$ &$[15.8, 16.8]$ &$[18.0, 19.0]$ &$2.20$ &...    &$0.15$ \smallskip \\
            &$F336W$, $F336W-K_{s}$ &...    &...    &...    &$[17.0, 18.0]$ &$0.05$ \smallskip \\
NGC\,2298   &$K_{s}$, $F606W-K_{s}$ &$[17.5, 18.5]$ &$[20.0, 21.0]$ &$2.50$ &...    &$0.10$ \smallskip \\
            &$F336W$, $F336W-K_{s}$ &...    &...    &...    &$[18.0, 19.0]$ &$0.035$\\ \midrule 
\enddata
\tablecomments{Locations of the most diagnostic regimes are defined as the region of minimum curvature (i.e.~second derivative crosses zero) in the case of the MS-saddle, and the bluest point on the MS, in the case of the MSTO.~The location of the SGB is identified empirically as the flattest point in the UV-near-IR CMDs.}
\end{deluxetable*}

\subsubsection{Goodness of Isochrone Fit}\label{sec:gof}
We defined a measure of the quality of the isochrone fit by minimizing the offset ``$D$" in color-magnitude space between the isochrone and the fiducial ridgeline as, 
\begin{equation} \label{eqn:d2}
D^{2} = \frac{1}{N-f}\sum_i^N\frac{(c_{\text{i,fid}}-c_{\text{i,iso}})^{2} + (m_{\text{i,fid}} - m_{\text{i,iso}})^{2}}{\sigma(c_{\text{i,fid}})^2 + \sigma(m_{\text{i,fid}})^2}
\end{equation} 
where $c_i$ is defined as the color, i.e.~$m_{\rm F606W} - m_{\text{K}_{\text{s}}}$ and $m_{\rm F336W} - m_{\text{K}_{\text{s}}}$, with uncertainty $\sigma(c_{\text{i,fid}})$ and $m_i$ the magnitude of each star with uncertainty $\sigma(m_{\text{i,fid}})$, summing over all $N$ stars.~The parameter $f$ is the number of degrees of freedom, or the number of parameters for which we are solving, which is three, for mean reddening, distance, and age.~This quantity was measured for each point along the isochrone, for both the optical-near-IR and UV-near-IR CMDs, for the case of different values of reddening and distance, letting both parameters wander away from the literature values.~The minimization of this quantity was then chosen as the desired method for which to determine the absolute ages of the clusters.~Additionally, a weighting function, to be described in the following section, was introduced in the calculation of $D^{2}$ along the characteristic color-magnitude scale length, $\Delta(\text{MSTO} - \text{MS-saddle})$, and in the regime of the SGB in the case of the UV-near-IR color combination.~The process of $D^{2}$ minimization was then repeated for each cluster six times, i.e.~for each isochrone set (\texttt{DSED}, \texttt{VR} and \texttt{BaSTI}) in two different color combinations $K_{s}$ vs.~F606W$-K_{s}$ and F336W vs.~F336W$-K_{s}$ in order to generate six independent determinations of the reddening, distance, and age, associated with each cluster.

The filter-set choice of this method was based upon the consideration to anchor the isochrone in a regime independent of reddening and distance, i.e. imaging both the MSTO and MSK, in the case of the $K_{s}$ vs.~F606W$-K_{s}$ CMD, and to explore the formally most diagnostic region associated with age determination, i.e.~the SGB, in the case for the F336W vs.~F336W$-K_{s}$ CMD.~This is illustrated in Figure~\ref{fig:all_cmds}, where we show the morphologies of these two CMDs in comparison with the pure $K_s$ vs.~$J\!-\!K_s$ CMD, which shows less overall diagnostic power as a result of the lower quality of the combined photometry.~We point out, however, that with deeper near-IR photometry the $K_s$ vs.~$J\!-\!K_s$ CMD is equally or even more diagnostic when compared with the other color combinations, especially at luminosities lower than the MSK.~Our preliminary tests show that the MS fainter than the MSK is sensitive to the chemical makeup of the stellar population, i.e.~its metal abundance and various element abundance ratios, including light and $\alpha$-elements.~This will be addressed in future papers of this series and is particularly important in light of upcoming third-generation telescopes, like JWST and ELT, which will be operating in this wavelength range.

To better leverage the diagnostic power of the MSTO-MS-saddle color-magnitude characteristic scale length, a weighting function was introduced in the $K_{s}$ vs.~F606W$-K_{s}$ isochrone fitting process of the entire MS.~This is also done for the case of the F336W vs.~F336W$-K_{s}$ CMD, with the weighting function designed to exploit the diagnostic power of the SGB.~In both cases the weighting function was subject only to the constraint that a maximum weight of 1.0 was achieved at the location of either the MS-saddle, MSTO or SGB, with the weights scaling as a function of distance in color-magnitude space from these points of reference.~Additionally, a multiplicative constant was introduced prior to the square root term as seen in Table~\ref{tab:msto_loc} in order to ensure that none of the determined weights had a value less than zero.~The weight function, designed purely to satisfy the given constraints, thus had the form
\begin{equation}
w=1 - a\sqrt{(m_{\text{DR1}} - m_{\text{data}})^{2}\times (m_{\text{DR2}} - m_{\text{data}})^{2}}, 
\end{equation}
where ``DR1,2" refers to the magnitude associated with the chosen diagnostic regime, i.e.~MSTO, MS-saddle and SGB. In the case of the F336W vs.~F336W$-K_{s}$ CMD, the MS-saddle-equivalent is chosen to be a point on the lower MS at ${\rm F336W}\!\approx\!22$\,mag.~The locations of the MSTO, MS-saddle and SGB as they appear in the two chosen color combinations are listed in Table~\ref{tab:msto_loc}, along with the aforementioned weighting coefficients.~Caveats associated with this choice of method including, the handling of photometric errors, and model limitations will be discussed in subsequent sections.

\begin{figure*}
	\centering
	\includegraphics[width=0.8\linewidth]{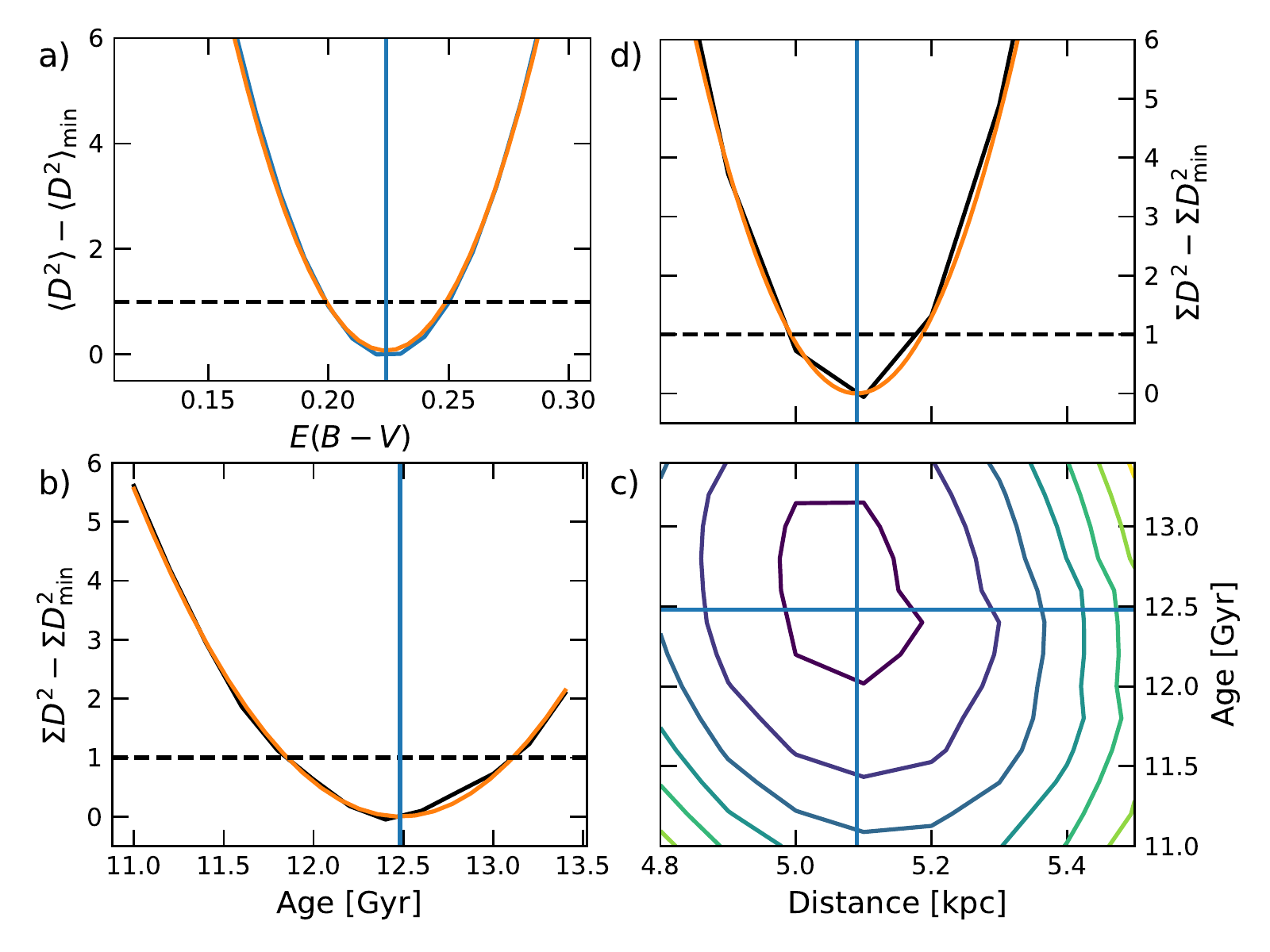}
	\caption{An example of the process for determining the mean reddening values, distances, and ages, with uncertainties, from Equation~\ref{eqn:d2} using \texttt{DSED} model comparisons to the NGC\,3201 $K_{s}$ vs.~F606W$-K_{s}$ photometry.~a) Mean $D^2$ over all distances and ages as a function of mean $E_{B-V}$.~The minimum of a second-order polynomial fit gives the most likely value of $E_{B-V}$.~The 1$\sigma$ uncertainty is taken where $\langle{D^2}\rangle_{\text{min}} + 1$ (horizontal dashed line).~b) The projection of $D^2$ over all distances as a function of age, the polynomial fit gives the best age and uncertainty.~c) Contours of $D^2$ over the distance and age parameter space with lines indicating the best-fit values of each parameter.~d) The projection of $D^2$ over all ages to determine the distance.}
	\label{fig:ebvdage}
\end{figure*}

\begin{deluxetable*}{ccccccc}
\tablewidth{0.85\textwidth}
\tablecaption{Results of the $D^{2}$ minimization procedure described in Section~\ref{txt:isofit}.~Errors attached to the determined values are derived as described in Section \ref{sec:gof}.\label{tab:chi2}}
\tablehead{\colhead{Cluster}    &\colhead{Model}    &\colhead{CMD}  &\colhead{Distance [kpc]}   &\colhead{$E_{B-V}$ [mag]}  &\colhead{Age [Gyr]}}
\startdata
NGC\,3201   & \texttt{\texttt{DSED}}        & $K_{s}$, $F606W-K_{s}$    & $5.1 \pm 0.1$    & $0.24 \pm 0.02$    & $12.0 \pm 0.7$ \\
            &               & $F336W$, $F336W-K_{s}$    & $5.0 \pm 0.2$    & $0.26 \pm 0.02$    & $12.4 \pm 0.8$ \\
            &               &                       &       &       &           \\
            & \texttt{\texttt{VR}}      & $K_{s}$, $F606W-K_{s}$    & $5.0 \pm 0.1$    & $0.22 \pm 0.03$    & $11.2 \pm 0.9$ \\
            &               & $F336W$, $F336W-K_{s}$    & $5.0 \pm 0.2$ & $0.26 \pm 0.03$   & $10.7 \pm 1.3$ \\
            &               &                       &       &       &           \\
            & \texttt{\texttt{BaSTI}}       & $K_{s}$, $F606W-K_{s}$    & $5.1 \pm 0.1$    & $0.21 \pm 0.02$    & $12.4 \pm 0.6$ \\
            &               & $F336W$, $F336W-K_{s}$    & $4.9 \pm 0.1$    & $0.26 \pm 0.02$    & $11.2 \pm 2.0$ \\ 
\midrule
NGC\,2298   & \texttt{\texttt{DSED}}        & $K_{s}$, $F606W-K_{s}$    &$10.7 \pm 0.3$    & $0.17 \pm 0.02$    & $13.1 \pm 0.6$ \\
            &               & $F336W$, $F336W-K_{s}$    & $10.6 \pm 0.2$   & $0.22 \pm 0.02$    & $13.2 \pm 0.6$ \\
            &               &                       &       &       &           \\
            & \texttt{\texttt{VR}}      & $K_{s}$, $F606W-K_{s}$    &$10.8 \pm 0.3$    & $0.13 \pm 0.03$    & $14.0 \pm 0.9$ \\
            &               & $F336W$, $F336W-K_{s}$    & $10.4 \pm 0.4$   & $0.21 \pm 0.02$    & $12.8 \pm 1.3$ \\
            &               &                       &       &       &           \\
            & \texttt{\texttt{BaSTI}}       & $K_{s}$, $F606W-K_{s}$    &$10.4 \pm 0.1$    & $0.14 \pm 0.01$    & $13.3 \pm 0.5$ \\
            &               & $F336W$, $F336W-K_{s}$    & $10.2 \pm 0.2$   & $0.22 \pm 0.02$   & $12.8 \pm 0.9$ \\ \midrule            
\enddata
\tablecomments{In the case of the \texttt{BaSTI} isochrones, conversions from the \cite{bes88} $K$ band to the 2MASS \cite{skrut06} system were made using the transformation equation provided by \cite{car01}: $K_{s,\text{2MASS}} = K_{BB} + (-0.039 \pm 0.007) + (0.001 \pm 0.005)\cdot(J-K)_{BB}$.}
\end{deluxetable*}

\begin{figure*}[ht!]
   \centering
   \includegraphics[width=0.87\linewidth]{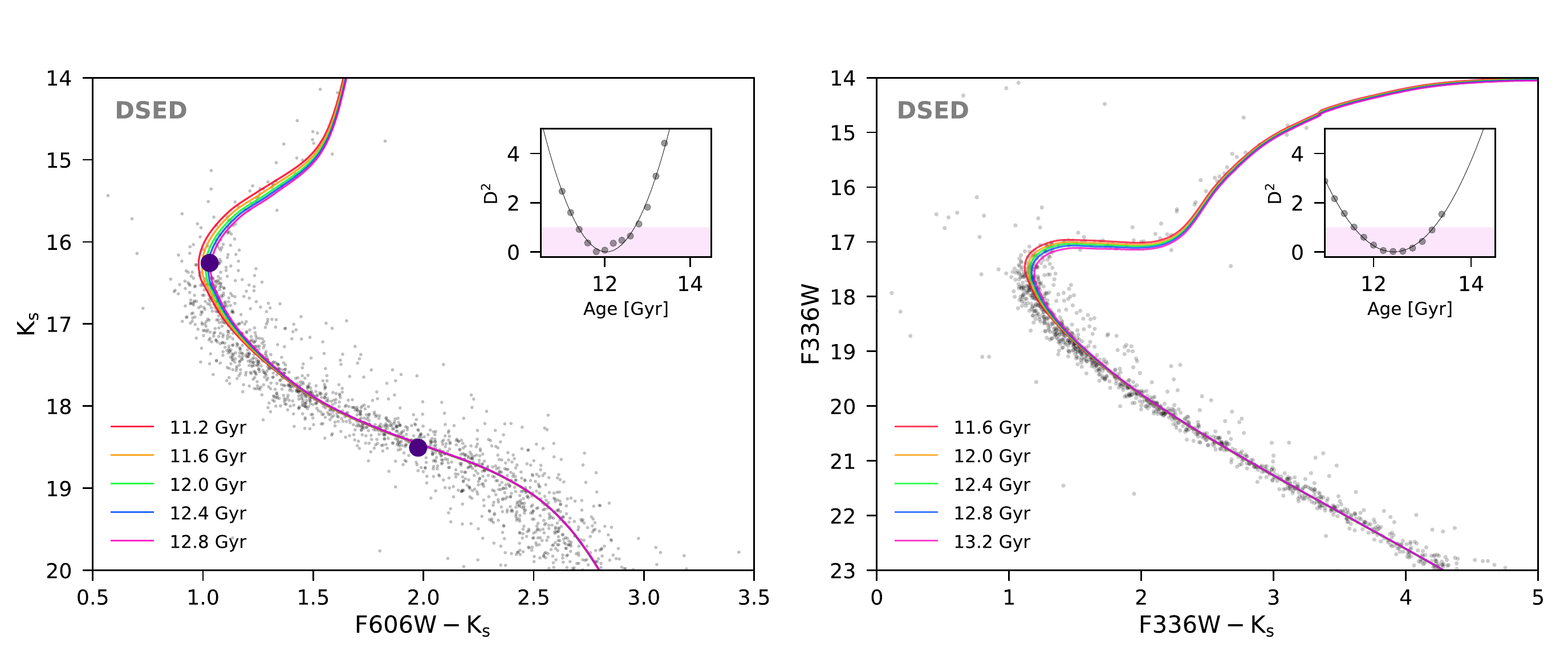}
   \includegraphics[width=0.87\linewidth]{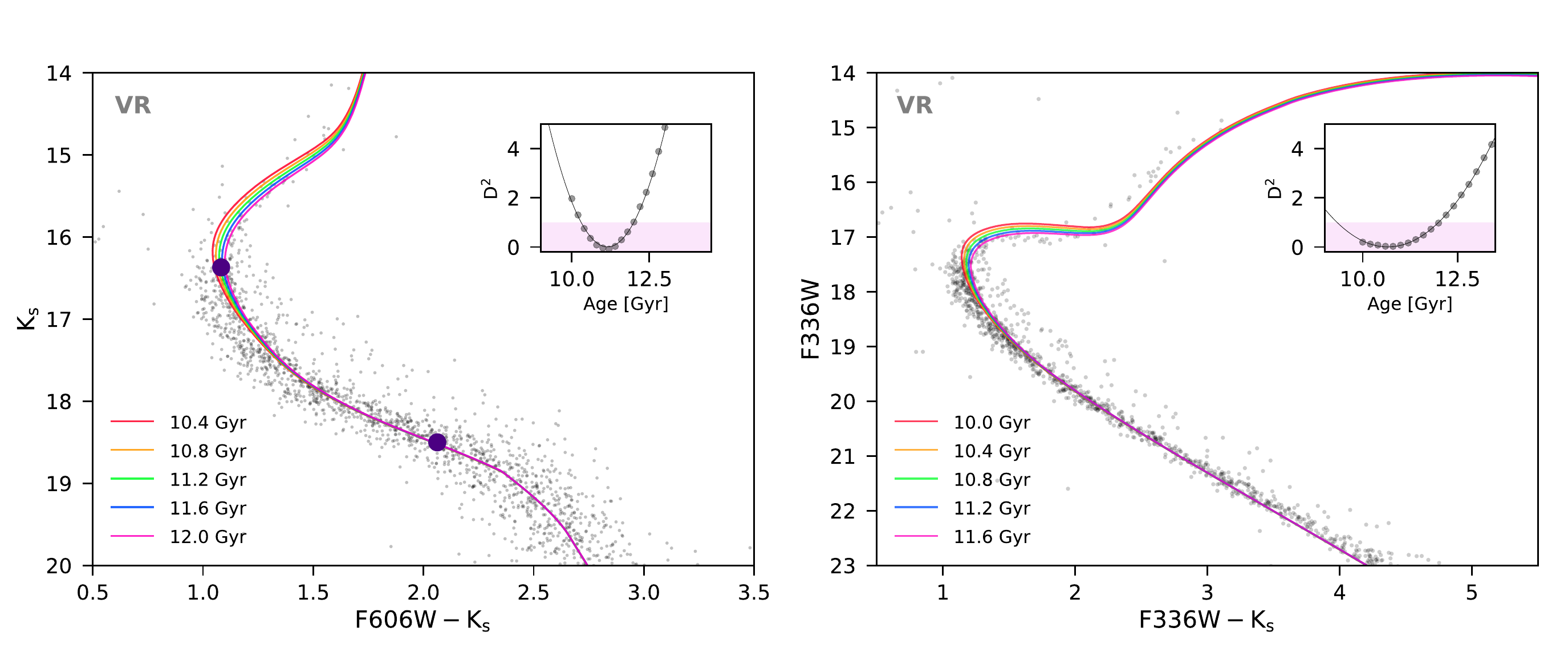}
   \includegraphics[width=0.87\linewidth]{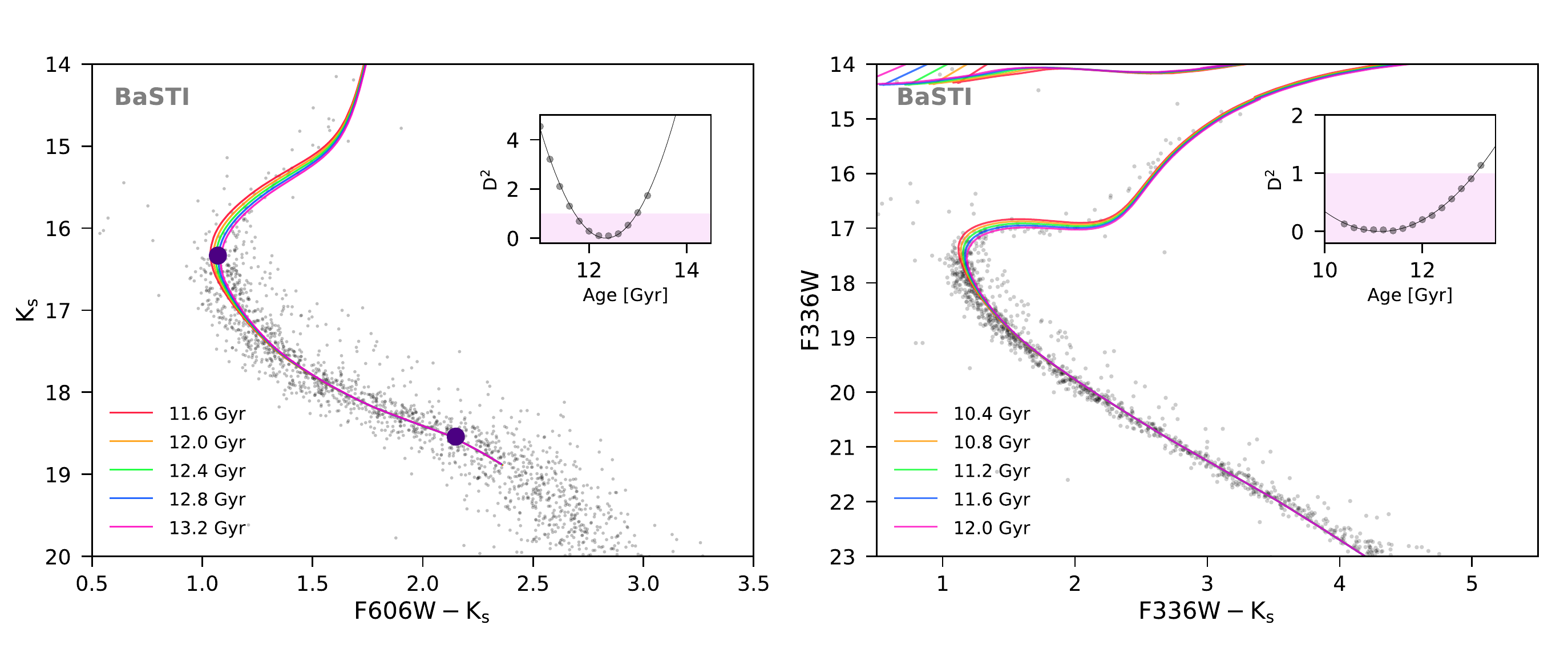}
   \caption{Summary of the $D^{2}$ minimization process for NGC\,3201 as described in Section~\ref{txt:isofit}.~({\it Top panels}): \texttt{DSED} isochrones, with fixed [Fe/H]=$-1.59$, [$\alpha$/Fe]=+0.2 and an age range of 10.0--14.0 Gyr for the optical-near-IR (left panel) and UV-near-IR CMDs (right panels).~({\it Middle panels}): \texttt{VR} isochrones with [Fe/H]=$-1.59$, [$\alpha$/Fe]=+0.2 and spanning 10.0--14.0 Gyr.~({\it Bottom panels}): \texttt{BaSTI} isochrones with fixed [Fe/H]=$-1.60$, [$\alpha$/Fe]=+0.2 and spanning 10.0--14.0 Gyr.~The insets included in all CMDs show the $D^{2}$ values for the various ages, with minimums given in Table \ref{tab:chi2}.~The red shaded area spans the lowest $D^{2}$ value to a value of 1.0 and defines the range of uncertainty in age.}
   \label{fig:3201_dse_chi}
\end{figure*}

\begin{figure*}[ht!]
   \centering
   \includegraphics[width=0.87\linewidth]{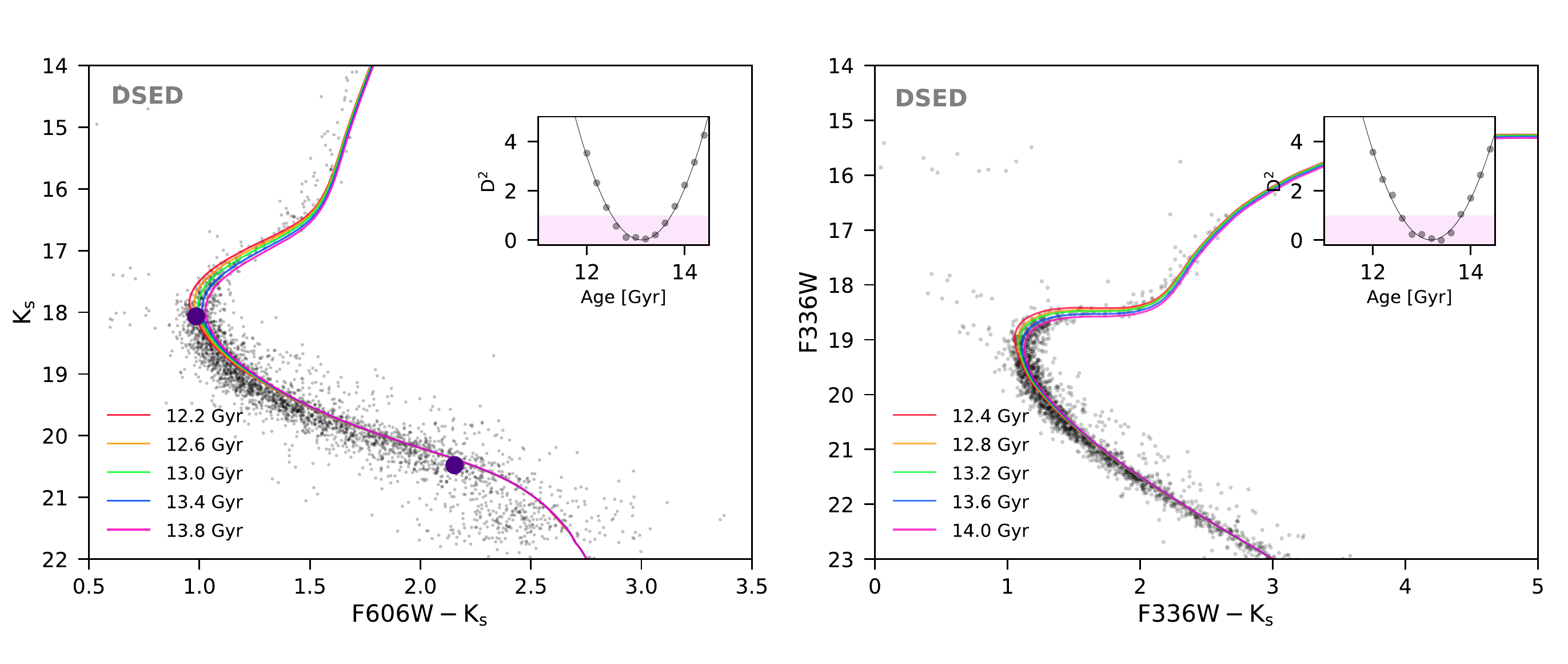}
   \includegraphics[width=0.87\linewidth]{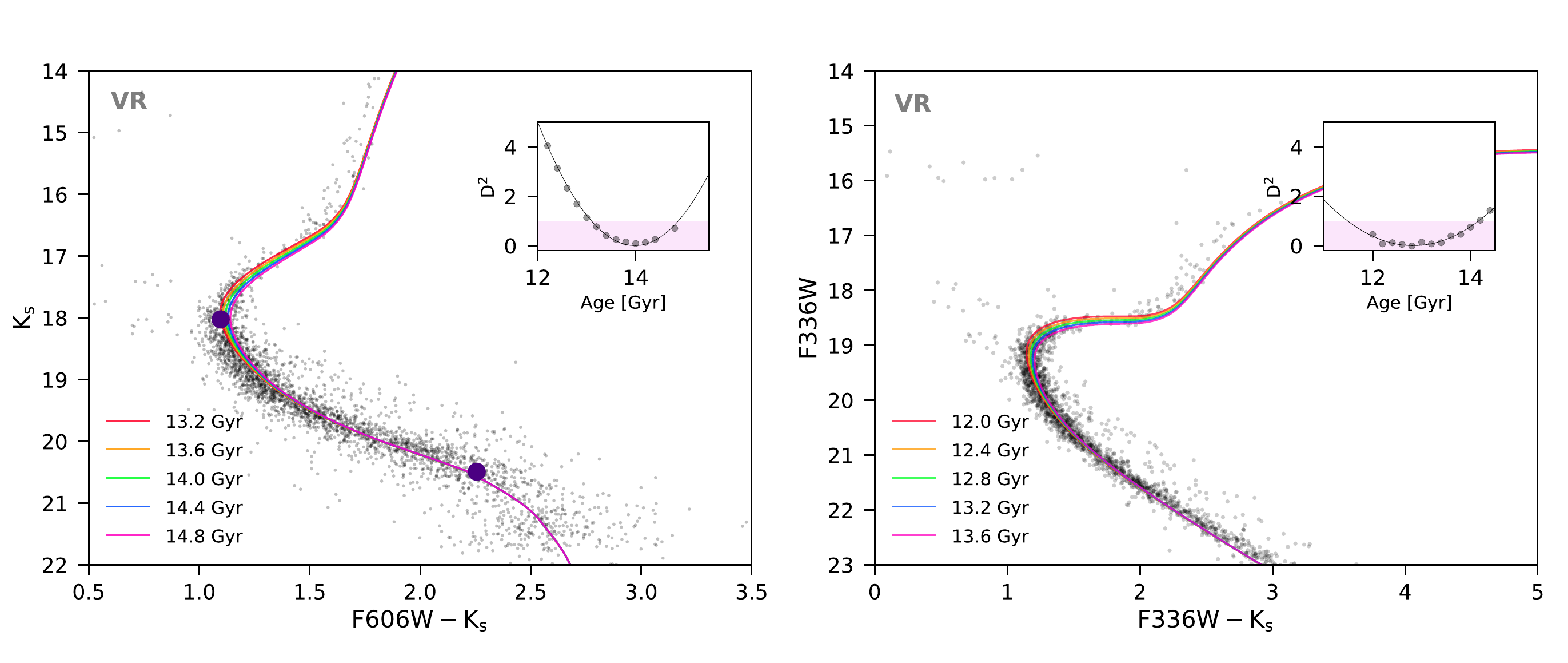}
   \includegraphics[width=0.87\linewidth]{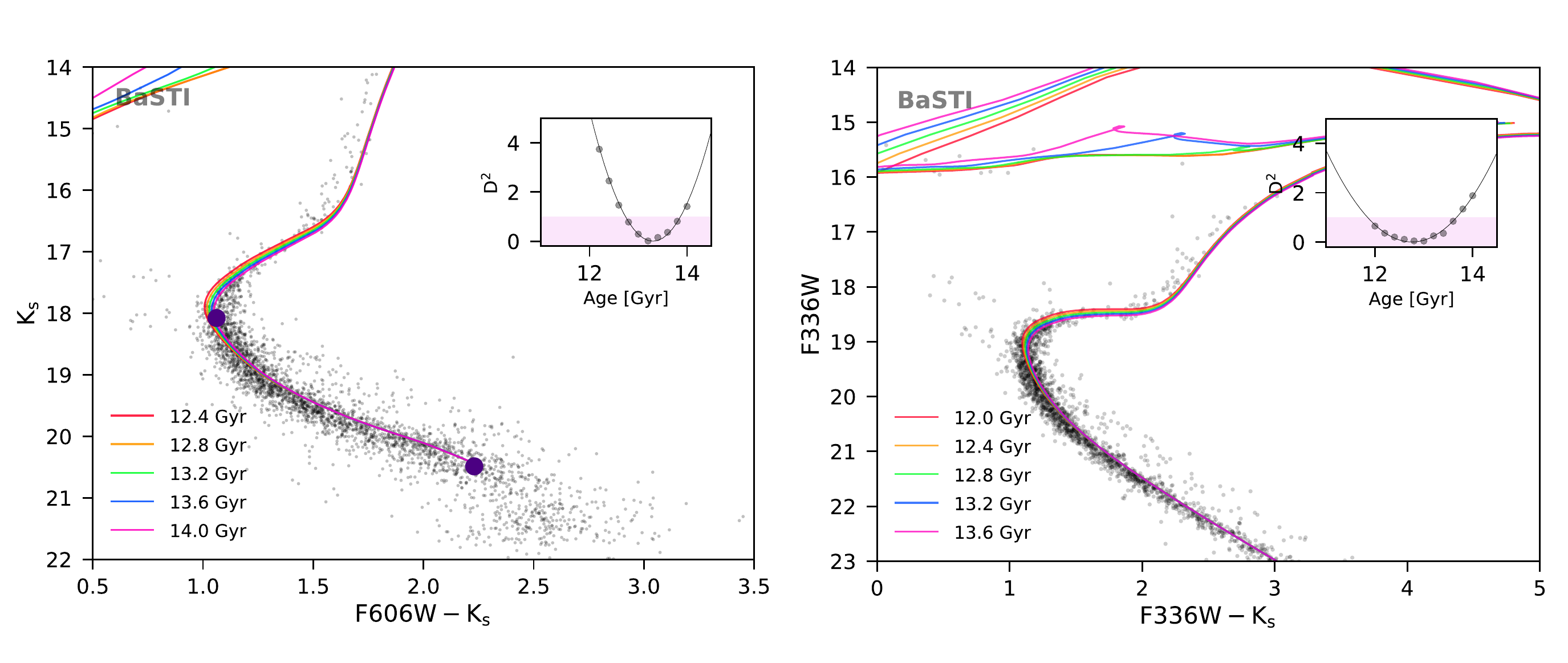}
   \caption{Summary of the $D^{2}$ minimization process for NGC\,2298 described in Section \ref{txt:isofit}.~({\it Top panels}): \texttt{DSED} isochrones, with fixed [Fe/H]=$-1.92$, [$\alpha$/Fe]=+0.2 and an age range of 12.0--15.0 Gyr for the optical-near-IR (left panel) and UV-near-IR CMDs (right panel).~({\it Middle panels}): \texttt{VR} isochrones with [Fe/H]=$-1.92$, [$\alpha$/Fe]=+0.2 and spanning 12.0--15.0 Gyr.~({\it Bottom panels}): Set of \texttt{BaSTI} isochrones with fixed [Fe/H]=$-1.90$, [$\alpha$/Fe]=+0.2 in the age range 12.0--15.0 Gyr.~The insets included in all CMDs show the $D^{2}$ values for the various ages, with minimums given in Table \ref{tab:chi2}.~The red shaded area spans the lowest $D^{2}$ value to a value of 1.0 and defines the range of uncertainty in age.}
   \label{fig:2298_dse_chi}
\end{figure*}
\begin{figure*}[ht!]
   \centering
   \includegraphics[width=0.85\linewidth]{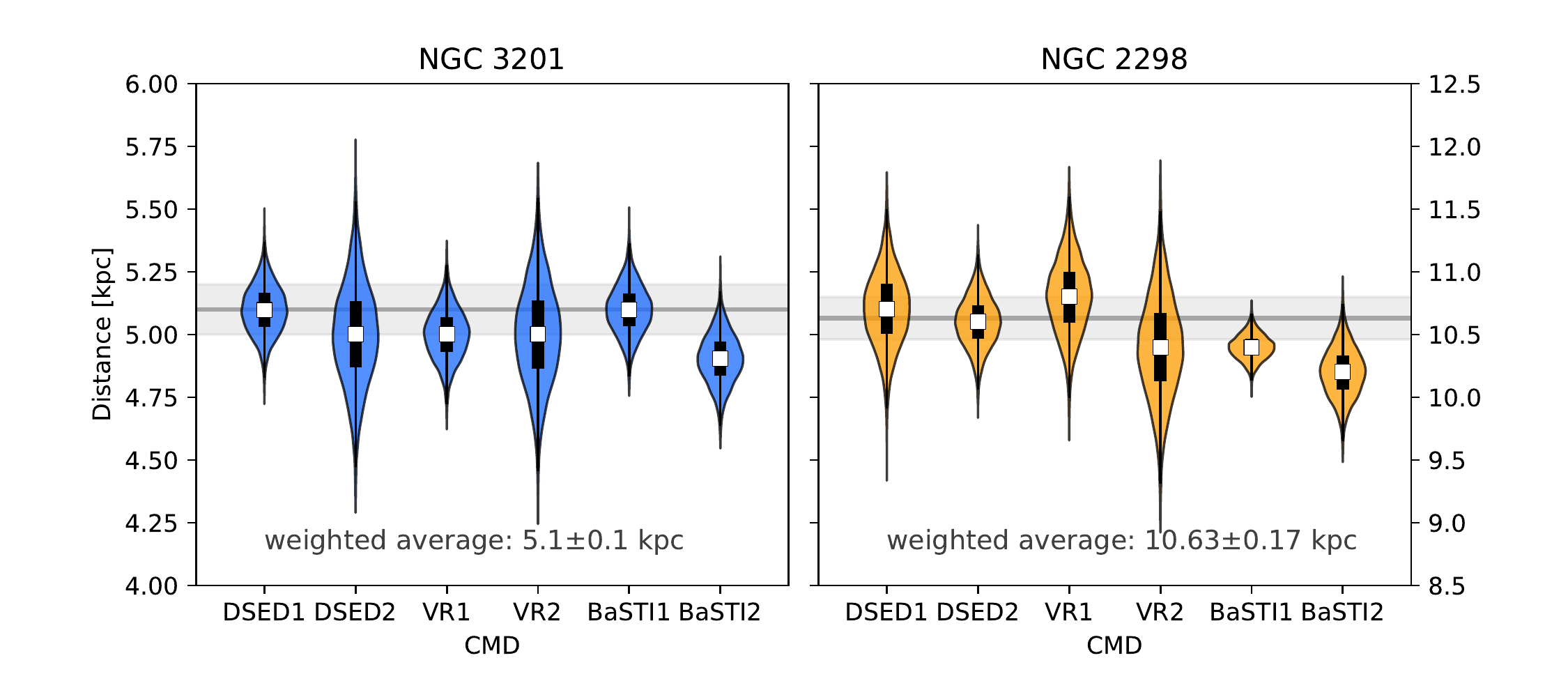}
   \includegraphics[width=0.85\linewidth]{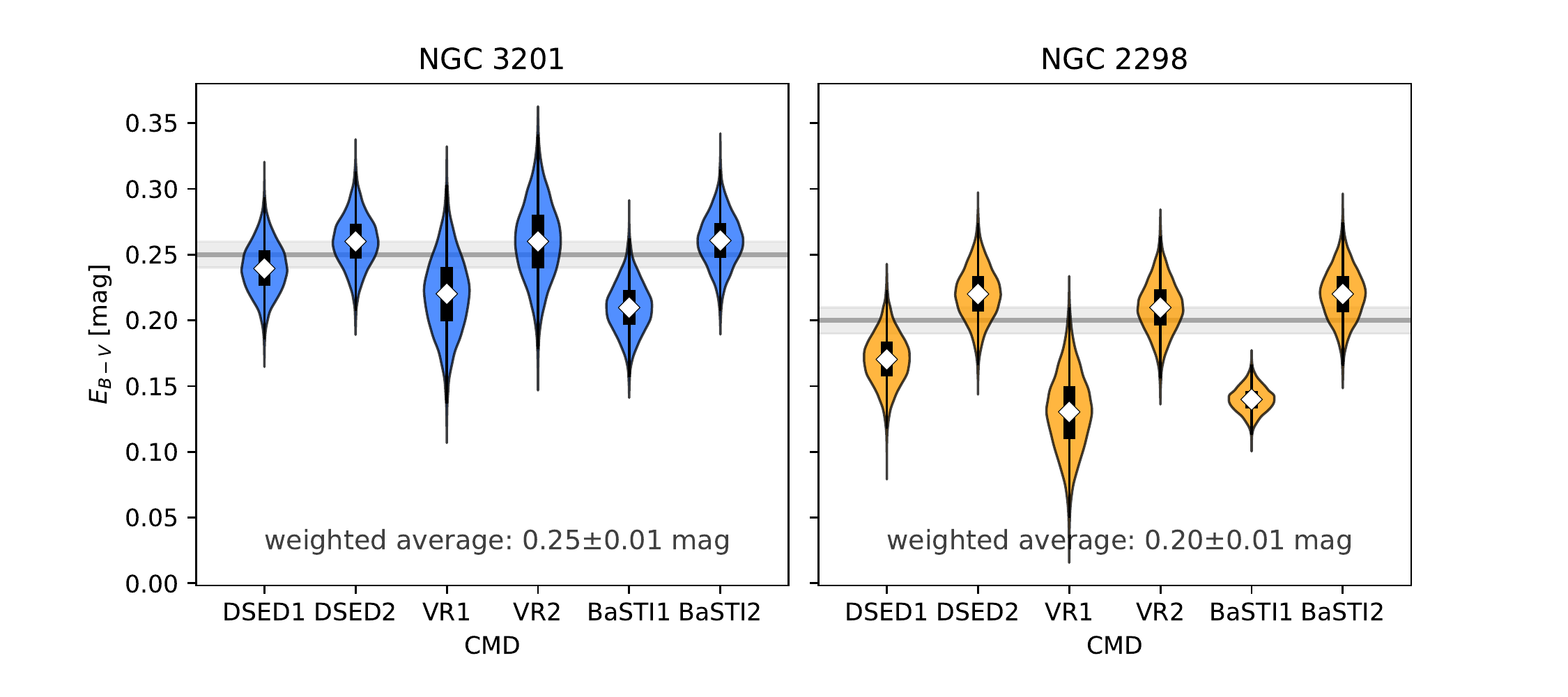}
   \includegraphics[width=0.85\linewidth]{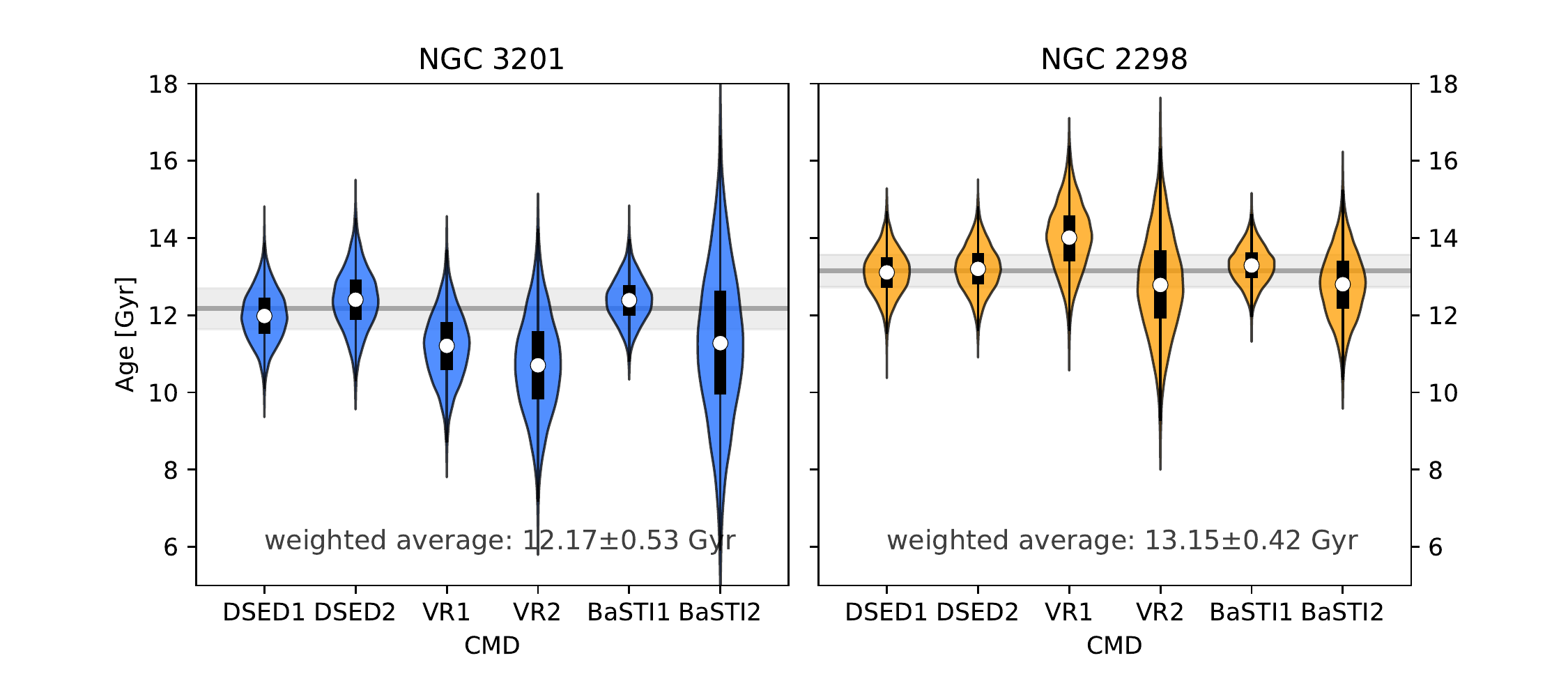}
   \caption{Violin plot comparison of our distance, reddening, and age determinations using the CMD combinations summarized in Table~\ref{tab:chi2} for NGC\,3201 (left panels) and NGC\,2298 (right panels).~The x-axis labels indicate the types of CMDs for each isochrone set, for example, DSED1 corresponds to the $K_{s}$ vs.~${\rm F606W}-K_{s}$ CMD, while DSED2 refers to the F336W vs.~${\rm F336W}-K_{s}$ CMD.~The same nomenclature is also used for the other two isochrone sets.~The horizontal lines and surrounding light-shaded regions represent the weighted average and error of the mean using the \texttt{DSED} isochrones, the numerical values of which are given in each panel.~Thick black vertical bars inside the violins indicate the $\pm1\sigma$ uncertainty ranges, while the violins are representing the corresponding Gaussian distributions.}
   \label{fig:violin1}
\end{figure*}

For each set of isochrones and filter combination, values of $D^2$ were calculated for each position in the mean reddening, distance, and age parameter space.~The choice to re-derive values of both reddening and distance was made in order to limit \textit{a priori} biasing of the age determination process through, for example, the introduction of an adjustment in color space in order to improve the overall isochrone fit, as has been done in previous studies.~A value of reddening was determined by first finding the mean $D^2$ for each $E_{B-V}$ over all distances and ages, $\langle{D^2}\rangle$.~A second order polynomial was then fit to $\langle{D^2}\rangle$ vs.\ $E_{B-V}$, with the minimum being the best-fit mean reddening.~The uncertainty is then $\Delta{E_{B-V}}$ from the point where $\langle{D^2}\rangle$ is $\langle{D^2}\rangle_{\text{min}} + 1$ (Figure~\ref{fig:ebvdage}a).~The distance and age were then determined for the best-fit mean $E_{B-V}$ by projecting the $D^2$ surface (Figure~\ref{fig:ebvdage}c).~The $D^2$ values were summed over all ages and the best-fit distance was the minimum of a second-order polynomial fit to $\Sigma{D^2}$ vs.\ distance.~Likewise, the best-fit age was derived from the projection of $D^2$ over all distances.~As with the reddening, the uncertainties in distance and age were determined where $\Sigma{D^2}\!=\!\Sigma{D^2}_{\text{min}}\!+\!1$.~Figure~\ref{fig:ebvdage} gives an example of the process.

The results of the $D^{2}$ minimization to determine reddening, distance, and absolute age for our two GCs are summarized in Table~\ref{tab:chi2}, illustrated in Figures~\ref{fig:3201_dse_chi} and \ref{fig:2298_dse_chi}, and discussed in the following sections.

\subsection{NGC\,3201}\label{sec:res_3201}
\subsubsection{Distance \& Reddening Comparison}\label{sec:3201distred}
Examining the results presented in Table \ref{tab:chi2}, we find good agreement regarding the derivations of distance and reddening both internally, within each isochrone set across the two color combinations and externally, comparing between different isochrone sets.~This consistency is illustrated in Figure~\ref{fig:violin1}, using violin plots.~We have chosen to select only the most \textit{internally} self-consistent isochrone set for final age determination, the \texttt{DSED} isochrones.~We provide in the each panel of Figure~\ref{fig:violin1} the corresponding weighted average values based on \texttt{DSED} isochrones only\footnote{We point out that averaging over one isochrone set avoids mixing different prescriptions in the underlying input physics used to generate the various isochrone models. However, the values for the other isochrone sets are provided in Table~\ref{tab:chi2} for transparency, should the reader wish to compute the global average.}.~The best-fit isochrone was found to be the [Fe/H]~$\!=\!-1.59$, [$\alpha$/Fe]~$\!=\!+0.2$, $Y\!=\!0.245$ from the \texttt{DSED} isochrone set because as previously stated, they provide the best internal consistency across distance, reddening, and age measurements and yield the smallest uncertainties.~Therefore, we compute the final values of reddening $E_{B-V}=0.25\pm0.01$ and distance $5.1\pm0.1$\,kpc using the weighted average of the two \texttt{DSED} isochrones (for the two color combinations).~It is these values that are discussed in the following in the context of previously derived literature values.

We obtain results that are in good agreement with previous studies, but note a number of discrepancies when comparing our best-fit values against individual works from the literature listed in Table~\ref{tab:comp3201} and illustrated in Figure~\ref{fig:violin2}.~We find the largest literature deviation of $\sim0.9$\,kpc to be between our NGC\,3201 best-fit distance ($5.1\pm0.1$\,kpc) and the work of of \cite{ang15}.~However, because of its unusually high uncertainty, the results of \cite{ang15} are still in agreement to within 2$\sigma$ of our determination, as is the case of the other literature values with uncertainty margins.~We do note, however, that overall our distance determination is slightly larger than all of the literature values.~Secondly, the best-fit reddening value ($E_{B-V}=0.25\pm0.01$) appears significantly lower than the value reported by \cite{pier02}, but again, is still within 2$\sigma$ agreement.~All other studies show similar, or better agreement, with our result.~Thirdly, we observe that our reddening values originating from the UV-near-IR CMDs are systematically larger than those from the optical-near-IR CMDs (see Fig.~\ref{fig:violin1}).

We stress that in our determination of reddening, particularly referencing the third point discussed above, we have chosen to fix our extinction coefficient, $R_{i}$, and adopt the literature value of $E_{B-V}$ as a starting point in our investigation.~Although we decided not to apply a shift in color-magnitude space prior to the age determination in order to achieve a better fit to our isochrones, we felt justified in exploring a range $E_{B-V}$ values and refer interested readers to \cite{case14} who caution against taking literature reddening values as absolute during isochrone fitting.~However, the authors also warn against the use of reddening as a completely free parameter, as it is only one of the many contributing factors to the quality of isochrone fits, and thus this is why we have adopted a literature value as a starting point.~In their work, \citeauthor{case14} investigated the effects of assuming a constant reddening law and extinction value for all spectral types in a CMD, concluding that it is necessary to either apply a shift in color-magnitude space to an isochrone to best approximate the MSTO, or to assume different extinction and reddening coefficients depending on the MSTO star type, differing by as much as $\Delta E_{B-V}\!=\!0.05$ and $\Delta R_{V}\!=\!0.3$.~As we have chosen not to apply such a shift, we conclude that our final deviation of up to $|\Delta E_{B-V}|\!\approx\!0.06$ from the literature value is within reason, given the wide SED coverage of our dataset.~In conclusion, we find that our distance and reddening measurements, subject to the discussed assumptions and uncertainties, range among the more statistically robust parameter determinations and are in good agreement with previously derived values (see Fig.~\ref{fig:violin2}).

\subsubsection{Age Comparison}
Subject to the discussed caveats associated with our method, and the internal accuracy of the isochrone sets in the regimes deemed most diagnostic for deriving ages, we now present a literature comparison of our final age determination for the cluster NGC\,3201, in addition to discussing the strengths and weaknesses of the various isochrone sets.~We determine an absolute best-fit age of $12.2\pm0.5$ Gyr, for NGC\,3201 using the weighted average from the two CMD types based on only the \texttt{DSED} isochrone set, for reasons discussed in the previous section.~Comparing our determined age with those listed in Table~\ref{tab:comp3201}, we see good agreement, within $2\sigma$ for most of the ages listed, including the age determined using the MSTO-MSK method derived by \cite{cal09}. However, we find a $4\!-\!5\sigma$ offset in the case of the age listed in \cite{roe14}, though note that the \cite{roe14} value is the youngest of the recent age determinations (see Figure~\ref{fig:violin2}).~Finally, we note that our age determination is among the best constrained age determination listed.

While our final age determination depends primarily on the optimization of the fit to the MSK, MSTO, and SGB, the quality of the fit to other regions of the CMD, in particular to the RGB in the case of the optical-near-IR color combination, degrades for all the isochrone sets.~Figure~\ref{fig:3201_dse_chi} shows that the isochrone sets consistently predict a redder RGB than is seen in the data.~This trend was also observed in earlier studies \citep[][and references therein]{sara16} and appears to be due to the isochrone model handling of the near-IR passbands.~In the case of the UV-near-IR CMD, the ill-fitted RGB again appears in all three isochrone sets, while the MS is fit best by the \texttt{DSED} isochrone, but is sufficiently well fit by all three isochrone sets.~However, the \texttt{BaSTI} isochrone fit to the UV colour combination only barely demonstrates the discovery of a clear minimum during isochrone fitting as can be seen in the inset of the UV CMD in the bottom-right panel of Figure \ref{fig:3201_dse_chi}.~This was also the case for the \texttt{VR} isochrone fit to the UV colour combination, resulting in large error bars for both the \texttt{BaSTI} and \texttt{VR} age determinations in the UV CMD.~Additionally, regarding the fit to the MSTO and MSK, shown as violet dots in Figure~\ref{fig:3201_dse_chi}, all three isochrone sets fit both regions well, with no obvious outliers.

\subsection{NGC\,2298}\label{sec:res_2298}
\subsubsection{Distance \& Reddening Comparison}
We list the results of isochrone fitting for NGC\,2298 in Table \ref{tab:chi2} and obtain the best-fit distance $10.6\pm0.2$\,kpc, again derived using the [Fe/H]~$\!=\!-1.92$, [$\alpha$/Fe]~$\!=\!+0.2$, $Y\!=\!0.245$ \texttt{DSED} isochrone set only, for reasons given in Section \ref{sec:3201distred}. This value is similar to previous determinations (see Fig.~\ref{fig:violin2}) by \cite{har10} and \cite{dut00}, but significantly smaller than the literature distance derived by \cite{pas09}.~Even our largest formal distance $10.8\pm0.3$\,kpc, derived for the \texttt{VR}-fit CMD is starkly inconsistent with the \cite{pas09} literature value.~Additionally, a trend can be observed for the best-fit reddening values, where the optical-near-IR values of $E_{B-V}$ appear systematically lower than from the UV-near-IR CMDs, similar to NGC\,3201 (see Fig.~\ref{fig:violin1}).~As discussed above, the reason could be related to changes in the reddening law towards both clusters which may be different from our assumed law (see Sect.~\ref{sec:syn_cat}).~We find a best-fit reddening towards NGC\,2298 of $E_{B-V}=0.20\pm0.01$, again adopting the weighted average from the \texttt{DSED} isochrones for comparison with the literature.~We find that our value is in very good agreement with previous measurements from \cite{omall17} and \cite{dut00}, but deviates substantially from \cite{har10}.~We propose that this disagreement could be a result of the methodology used by \cite{har10}, as integrated-light measurements may suffer from differential reddening issues (see Sect.~\ref{sec:dif_red}).~Although the cause of deviations between the literature and derived values of distance and reddening is not obvious in each case, or easily inferred for this cluster, we note that the deviations in distance are within the expected accuracy of the literature distance measurements \citep[see][]{bon08}.~Finally, we point out that our distance and reddening determinations mark the only statistically meaningful constraints for NGC\,2298 measured thus far (see Fig.~\ref{fig:violin2}).

\subsubsection{Age Comparison}
\begin{figure*}[!ht]
   \centering
   \includegraphics[width=0.85\linewidth]{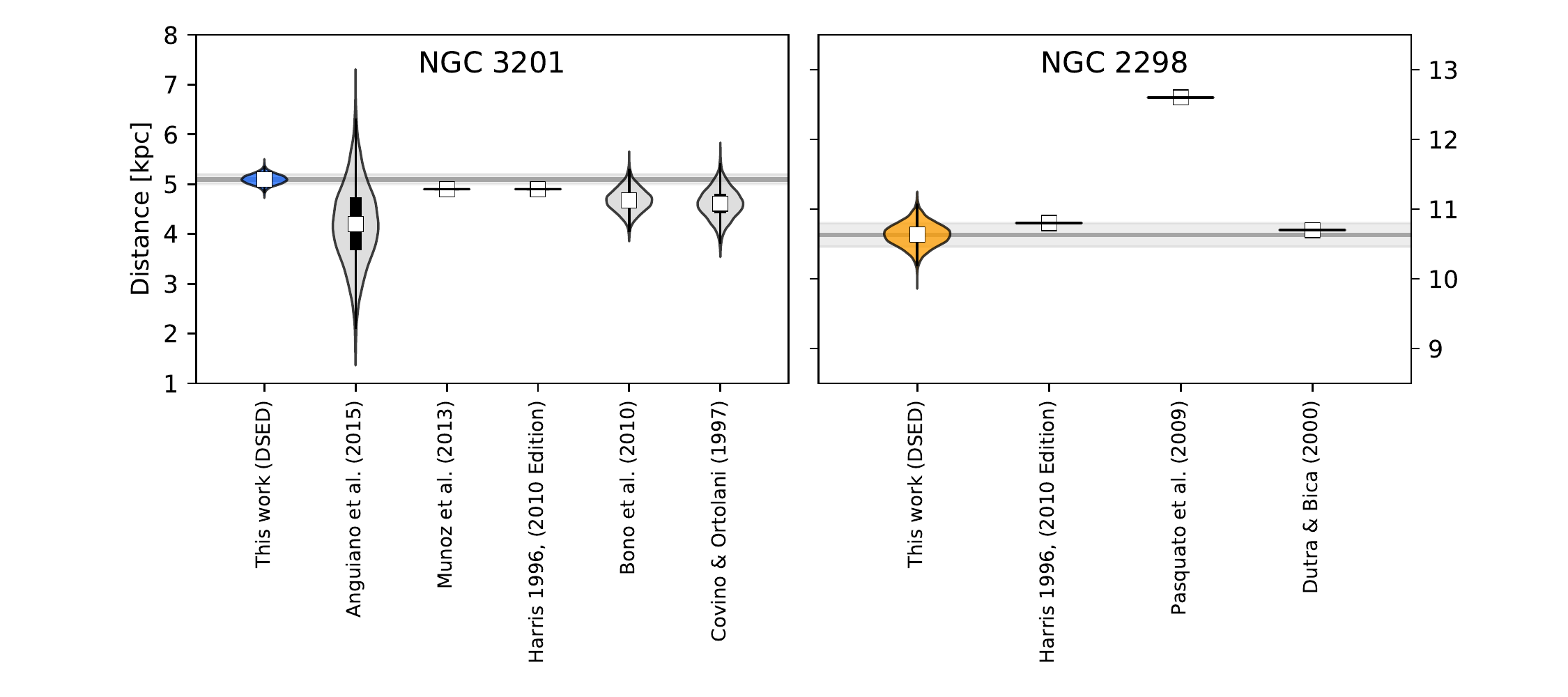}
   \includegraphics[width=0.85\linewidth]{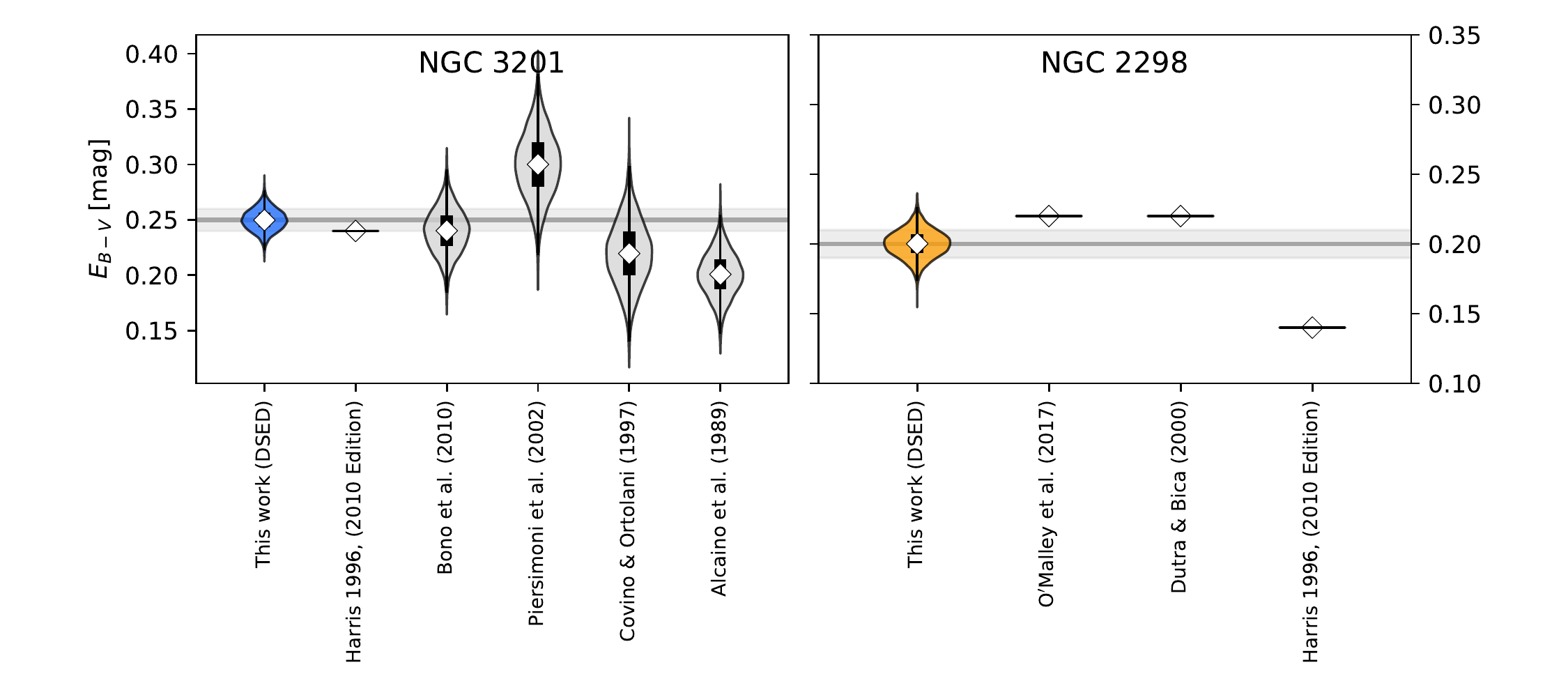}
   \includegraphics[width=0.85\linewidth]{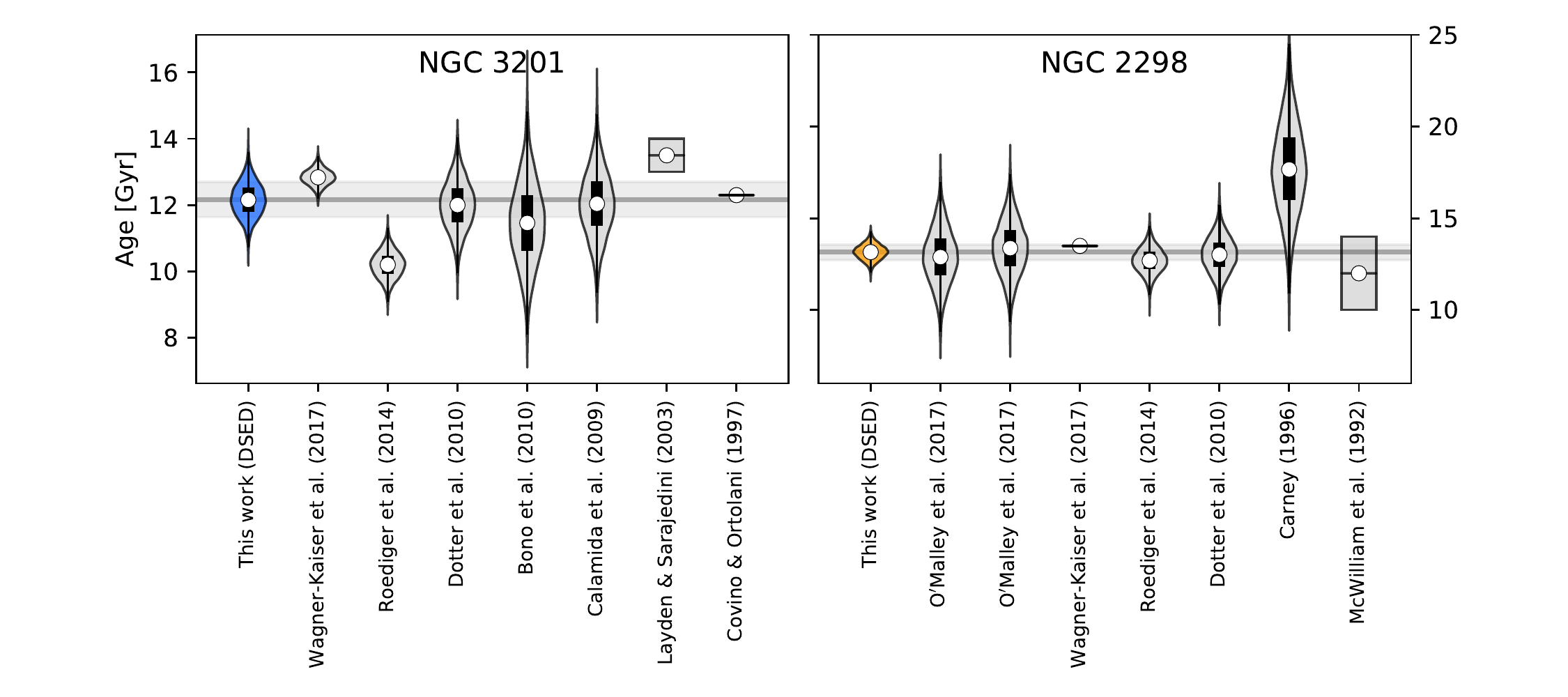}
   \caption{Violin plot comparisons of our distance, reddening, and age determinations with literature values for NGC\,3201 (left panels) and NGC\,2298 (right panels).~The x-axis labels refer to the publications listed in Tables~\ref{tab:comp3201} and \ref{tab:comp2298}.~The horizontal lines and surrounding light-shaded regions are as in Figure~\ref{fig:violin1}, and represent the weighted mean and error of the mean computed using the \texttt{DSED} isochrones only.~The violin labeled 'This work' also displays the value.~Vertical boxes indicate published ranges, while horizontal bars without vertical spread show literature values for which no uncertainties were given in the corresponding paper.~Thick black vertical bars inside the violins indicate the $\pm1\sigma$ uncertainty ranges, while the violins are representing the corresponding Gaussian distributions.}
   \label{fig:violin2}
\end{figure*}

We derive a best-fit age of $13.2\pm0.4$ Gyr for the cluster NGC\,2298, using the weighted \texttt{DSED} isochrone set, grouping it among the oldest MW GCs.~A comparison of the derived age against literature values is given in Table~\ref{tab:comp2298} and illustrated in Figure~\ref{fig:violin2}.~We find good agreement between our results and previous studies to within $\sim\!1\sigma$.~Although the literature value from \cite{car96} is $\sim\!5$ Gyr off compared to our value and the rest of the literature, this offset amounts to only $\sim\!2\sigma$.~Despite the fact that our determined age is among the oldest (modern) ages measured, we find it to be one of the best constrained CMD age measurements for that cluster found in the literature.~The only exception is the value reported by \cite{wag17} who derive an age of $13.493^{+0.007}_{-0.017}$~Gyr from a Bayesian analysis.~Given that we use the same ACS photometry as \cite{wag17}, the Bayesian uncertainties may be underestimated.

Examining the isochrone fits presented in Figures \ref{fig:2298_dse_chi}, we can see that the \texttt{VR} isochrone set reproduces the position of the MSK with the highest fidelity, while all three isochrone sets approximate the MSTO fairly well.~Additionally, we observe the same trend as was observed for NGC\,3201, namely, that all three isochrone sets predict a redder RGB than what is observed in the data.~The \texttt{VR} isochrone best models the data at luminosities below the MSK, with the \texttt{DSED} and \texttt{BaSTI} isochrone sets either laying above the majority of the data points below the MSK or terminating before a trend is established (in the case of the \texttt{BaSTI} isochrone).~Considering the fit of the isochrones to the UV-near-IR CMD, all three isochrone sets demonstrate good overall fits from the MS through the RGB.~However, all three isochrone sets also appear bluer than the bulk of the data at the MSTO, with the data appearing to ``puff up" as the MS transitions to the SGB.~This broadening could be the result of uncorrected DR or the result of the presence of multiple stellar populations in this cluster \citep[see Figure 4 in][]{mil12b}.

\begin{deluxetable*}{lcccc}
\tablewidth{0.85\textwidth}
\tablecaption{Comparison of the derived NGC\,3201 parameters with the literature. \label{tab:comp3201}}
\tablehead{\colhead{Parameter} &\colhead{Value}         &\colhead{Source}       &\colhead{Deviation}}
\startdata
Distance [kpc]  &$4.2\pm0.8$                &\cite{ang15}           &$+0.9$ \\
                &$4.9$                      &\cite{har10, mun13}    &$+0.2$ \\
                &$4.68\pm0.24$              &\cite{bon10}           &$+0.42$ \\
                &$4.6\pm0.3$                    &\cite{cov97}           &$+0.5$ \smallskip \\      
$E_{(B-V)}$     &$0.24$                     &\cite{har10}           &$+0.01$ \\
                &$0.24\pm0.02$              &\cite{bon10}           &$+0.01$ \\
                &$0.30\pm0.03$              &\cite{pier02}          &$-0.05$ \\
                &$0.22\pm0.03$              &\cite{cov97}           &$+0.03$ \\
                &$0.20\pm0.02$              &\cite{alc89}           &$+0.05$ \smallskip \\
Age [Gyr]           &$12.836^{+0.277}_{-0.206}$     &\cite{wag17}           &$-0.666$    \\
                &$10.2\pm0.4$               &\cite{roe14}           &$+1.97$ \\
                &$12.0\pm0.75$              &\cite{dot10}           &$+0.17$ \\
                &$11.48\pm1.27$             &\cite{bon10}           &$+0.67$ \\
                &$12.0\pm1.0$               &\cite{cal09}           &$+0.17$ \\
                &$13.0 - 14.0$              &\cite{lay03}           &$-0.83$ \\
                &$12.3$                     &\cite{cov97}           &$-0.13$  \\
\enddata
\tablecomments{The quoted deviation is in reference to the corresponding values computed using the weighted average of the two CMD types from the \texttt{DSED} isochrones only.}
\end{deluxetable*}

\begin{deluxetable*}{lcccc}
\tablewidth{0.85\textwidth}
\tablecaption{Comparison of the derived NGC\,2298 parameters with the literature. \label{tab:comp2298}}
\tablehead{\colhead{Parameter} &\colhead{Value}         &\colhead{Source}       &\colhead{Deviation} }
\startdata
Distance [kpc]  &$10.8$                     &\cite{har10}           &$-0.20$ \\
                &$12.6$                     &\cite{pas09}           &$-2.0$ \\
                &$10.7$                     &\cite{dut00}           &$-0.1$ \smallskip \\
$E_{(B-V)}$     &$0.22$                     &\cite{omall17, dut00}  &$-0.02$ \\
                &$0.14$                     &\cite{har10}           &$+0.06$ \smallskip \\ 
Age [Gyr]       &$12.9\pm1.5$               &\cite{omall17}         &$+0.25$ \\
                &$13.4\pm1.5$               &\cite{omall17}         &$-0.25$ \\
                &$13.493^{+0.007}_{-0.017}$ &\cite{wag17}           &$-0.343$ \\
                &$12.7\pm0.7$               &\cite{roe14}           &$+0.45$ \\
                &$13.0\pm1.0$               &\cite{dot10}           &$+0.15$ \\
                &$17.7\pm2.5$               &\cite{car96}           &$-4.55$ \\
                &$10 - 14$                  &\cite{mcwill92}        &$+1.15$ \\
\enddata
\tablecomments{The quoted deviation is in reference to the corresponding values computed using the weighted average of the two CMD types from the \texttt{DSED} isochrones only.}
\end{deluxetable*}

%%%%%%%%%%%%%%%%%%%%%%%%%%%%%%%%%%%%%%%%
\section{Summary \& Conclusions}
\label{sec:sum}
The multi-conjugate adaptive optics system, near-IR imager combination, GeMS/GSAOI, mounted on the 8.1-m Gemini-South telescope was used to observe two Milky Way globular clusters, NGC\,3201 and NGC\,2298, reaching a depth of $K_s\!\simeq\!21$ Vega mag for both targets.~Spatially variable PSFs were created for both clusters using \texttt{DAOPHOT-IV} in order to perform PSF-fitting photometry, yielding high-quality photometric results.~The resulting photometric catalogues were combined with HST-near-UV data from \cite{pio15} and HST-optical data from \cite{sar07} to create panchromatic stellar photometry libraries for both clusters spanning near-UV to near-IR wavelengths.~Zero-point calibrations, proper motion cleaning and differential reddening corrections were applied to said catalogues to provide clean, precise photometry and to ensure that they contained only high-probability GC member stars.~Absolute age determinations were then performed, utilizing a characteristic scale length defined as the distance from MS-saddle to MSTO for both clusters.~The lower MSK was recovered in both clusters, for the first time in NGC\,2298, highlighting the diagnostic power of AO-supported near-IR data.~Three different isochrone sets (\texttt{DSED}, \texttt{VR} and \texttt{BaSTI}) were used in combination with two CMDs created using UV-near-IR (F336W vs.~F336W$-K_{s}$) and optical-near-IR ($K_s$ vs.~F606W$-K_{s}$) filter combinations to derive values of absolute age, distance and reddening, adopting values of [Fe/H] and [$\alpha$/Fe] derived via high-resolution spectroscopy of member stars.~The most internally consistent results were determined using the \texttt{DSED} isochrone-based measurements, yielding an age, reddening, and distance determination for NGC\,3201 of $12.2\pm0.5$ Gyr, $E_{B-V}=0.25\pm0.01$ mag, and $5.1\pm0.1$ kpc, respectively.~The \texttt{DSED} isochrones were also found to be the most internally consistent set for NGC\,2298, yielding age, reddening and distance determinations of $13.2\pm0.4$ Gyr, $E_{B-V}=0.20\pm0.01$ mag, and $10.6\pm0.2$ kpc, respectively.~We find very good agreement between literature values and our derived parameters, and show that our measurements are among the most statistically robust constraints determined thus far.~New observations for the G4CS are planned in order to apply this method to globular clusters with a wider range of metallicities and formation histories.

\acknowledgments
We thank Giampaolo Piotto for kindly providing us with the intermediate data release photometry from the HST Treasury Program GO-13297 \citep{pio15}.~We thank the anonymous referee for providing constructive remarks that sharpened the results presented in this paper.~S.~Monty would like to thank her collaborators, Gemini Observatory, the Institute of Astrophysics at the Pontificia Universidad Cat\'olica de Chile, R.~Monty Bird, and the University of Victoria for their support and hospitality and for the incredible learning opportunity provided through the University of Victoria Physics and Astronomy Co-op program.~T.H.P.\ acknowledges support through the FONDECYT Regular Project Grant No.~1161817 and the BASAL Center for Astrophysics and Associated Technologies (PFB-06).

This research is based on observations obtained at the Gemini Observatory, which is operated by the Association of Universities for Research in Astronomy, Inc., under a cooperative agreement with the NSF on behalf of the Gemini partnership: the National Science Foundation (United States), the National Research Council (Canada), CONICYT (Chile), the Australian Research Council (Australia), Minist\'{e}rio da Ci\^{e}ncia, Tecnologia e Inova\c{c}\~{a}o (Brazil) and Ministerio de Ciencia, Tecnolog\'{i}a e Innovaci\'{o}n Productiva (Argentina). Additionally, this research has made use of the NASA/IPAC Extragalactic Database (NED) which is operated by the Jet Propulsion Laboratory, California Institute of Technology, under contract with the National Aeronautics and Space Administration.

\vspace{5mm}
\facilities{Gemini: South (GeMS/GSAOI); HST (ACS, WFC3)}

\software{\texttt{astropy} \citep{astropycol013},\, \texttt{matplotlib} \citep{hun07},\, \texttt{NumPy}\,\citep{vdq011},\, \texttt{\texttt{IRAF}},\, \texttt{THELI} \citep{erb05,sch13},\, \texttt{DAOPHOT} \citep{stet87}}

\end{document}